\documentclass[a4paper,12pt]{article}
\pdfoutput=1
\usepackage{jheppub}
\usepackage{graphicx}
\usepackage{amssymb}
\usepackage{amsmath}
\usepackage{amsfonts}
\usepackage{hyperref}
\usepackage{xcolor}

	\def\bea{\begin{eqnarray}}
	\def\eea{\end{eqnarray}}
	\def\be{\begin{equation}}
	\def\ee{\end{equation}}
	\def\nn{\nonumber\\}

	\def\mm{\mathcal}
	
	\def\pa{\partial}

	\definecolor{db}{rgb}{0,0.08,0.45}
	\definecolor{brick}{rgb}{0.6,0.1,0.3}
	
	\definecolor{zz}{rgb}{1,0,0}
	\definecolor{zz2}{rgb}{0.7,0.1,0.1}
	\definecolor{yy}{rgb}{0.05,0.9,0.05}
	\definecolor{ww}{rgb}{0.6,0.1,0.3}
	
	\definecolor{rr}{cmyk}{0,0,0,1}
	\definecolor{vv}{rgb}{0.5,0,0.5}
	\definecolor{ss}{cmyk}{0,0,0,1}
	
	\def\zz{\textcolor{zz}}

	\definecolor{brick}{rgb}{0.5,0,0.5}

	\def\a{\alpha} \def\b{\beta}  \def\d{\delta}    \def\n{\nu} \def\m{\mu}     \def\l{\lambda} \def\o{\omega} 
	\def\G{\Gamma}  \def\D{\Delta}      \def\O{\Omega}

\title{Loops in AdS: From the Spectral Representation to Position Space II}

\author[\Psi]{Dean Carmi}
\affiliation[\Psi]{Department of Mathematics and Physics University of Haifa at Oranim, Kiryat Tivon 36006, Israel}
\emailAdd{deancarmi1@gmail.com}

\abstract{
We continue the study of AdS loop amplitudes in the spectral representation and in position space. We compute the finite coupling 4-point function in position space for the large-$N$ conformal Gross Neveu model on $AdS_3$. The resummation of loop bubble diagrams gives a result proportional to a tree-level contact diagram.  We show that certain families of fermionic Witten diagrams can be easily computed from their companion scalar diagrams. Thus, many of the results and identities of \cite{Carmi:2019ocp} are extended to the case of external fermions. We derive a spectral representation for ladder diagrams in AdS. Finally, we compute various bulk 2-point correlators, extending the results of \cite{Carmi:2019ocp}.}

\begin{document} 
\maketitle
\flushbottom



\section{Introduction}
Two very important problems in physics are: understanding quantum gravity and understanding strongly coupled (non-perturbative) quantum field theories. These two problems are beautifully connected to each other via the gauge-gravity duality, in which Anti-de Sitter (AdS) space plays a unique role. Therefore computing observables for quantum field theories defined on AdS space gives important insights about the two questions mentioned above.
AdS space, being a maximally symmetric space-time, often enables to perform analytic computations of observables in the QFT on AdS. For a very partial list of references see  \cite{Maldacena:1997re,Gubser:1998bc,Witten:1998qj,Callan:1989em,Breitenlohner:1982jf,Aharony:1999ti,Aharony:2010ay,Aharony:2012jf,Burges:1985qq,Witten:1998zw,Burgess:1984ti,Inami:1985dj}. 

An important observable for a quantum field theory defined on AdS space is the scattering amplitude. The amplitude can be computed perturbatively, similarly to Feynman diagrams in flat space. Through the AdS/CFT correspondence, AdS scattering amplitudes compute CFT correlation functions on the boundary of AdS.
Computations of AdS scattering amplitudes have been performed in position-space \cite{Liu:1998th,Liu:1998ty,Dolan:2000ut,Freedman:1998tz,DHoker:1998ecp,Freedman:1998bj,DHoker:1998bqu,DHoker:1999mqo,Zhou:2018sfz,Hijano:2015zsa}, in Mellin-space \cite{Penedones:2010ue,Fitzpatrick:2011ia,Paulos:2011ie,Rastelli:2017udc,Rastelli:2016nze,Cardona:2017tsw,Yuan:2017vgp,Yuan:2018qva}, momentum-space \cite{Raju:2010by,Raju:2011mp,Albayrak:2019asr,Albayrak:2018tam,Albayrak:2020fyp,Albayrak:2020isk,Armstrong:2020woi}, embedding space \cite{Costa:2014kfa,Costa:2011mg}. Most of these computations are restricted to tree-level amplitudes. Other computations rely on the conformal symmetry or the supersymmetry of the dual theory \cite{Aharony:2016dwx,Henriksson:2017eej,Alday:2017xua,Alday:2017gde,Mazac:2018mdx,Carmi:2020ekr,Caron-Huot:2018kta,Alday:2017vkk,Aprile:2017bgs,Aprile:2017qoy,Bissi:2020woe,Bissi:2020wtv,Ferrero:2021bsb,Ferrero:2019luz}. For additional works on AdS amplitudes and in particular loop amplitudes, see \cite{Carmi:2018qzm,Fitzpatrick:2010zm,Fitzpatrick:2011hu,Bertan:2018khc,Bertan:2018afl,Beccaria:2019stp,Fitzpatrick:2011dm,Ponomarev:2019ofr,Giombi:2017hpr,Costantino:2020vdu,Antunes:2020pof,Meltzer:2019nbs,Meltzer:2020qbr,Albayrak:2020bso,Nagaraj:2020sji,Zhou:2020ptb,Eberhardt:2020ewh}.

In this work we continue the study of loop Witten diagrams initiated in \cite{Carmi:2019ocp}, and extend those results in several directions. In section~\ref{sec:relswitten1} we consider Witten diagrams with external fermions. Using an identity between bulk-to-boundary fermion and scalar propagators, we show that certain families of fermionic diagrams can be directly computed from similar diagrams with external scalars. In section~\ref{sec:grossneveu} we compute the finite coupling 4-point function of the conformal large-$N$ Gross-Neveu model on $AdS_3$. The result of an infinite sum of loop bubble diagrams is extremely simple: it is proportional to a tree-level contact diagram 4-point function. In section~\ref{sec:ladder} we consider scalar ladder diagrams in AdS with an arbitrary number of rungs. We derive a spectral representation for ladder diagrams in AdS. In section~\ref{sec:2point6} we compute bulk-to-bulk 2-point correlators in several cases: scalar sunset diagrams in $AdS_3$, and bubble diagrams for scalars in $AdS_5$ and for fermions in $AdS_3$. In Appendix~\ref{sec:AA2} we show the details of the calculation of the spectral representation of the 1-loop box diagram. In Appendix~\ref{sec:b2} we speculate about an eigenvalue equation for ladder diagrams in AdS. In Appendix~\ref{sec:c2} we compute a 4-point correlator for the $O(N)$ model on $AdS_5$. In Appendix~\ref{sec:5} we discuss 4-point bubble diagrams for a scalar $\phi^4$ theory on $AdS_5$. In Appendix~\ref{sec:E2} we discuss 4-point bubble diagrams for a scalar $\phi^4$ theory on $AdS_2$.



\section{Witten diagrams with external Fermions}
\label{sec:relswitten1}

\begin{figure}[t]
\centering
\includegraphics[clip,height=3.3cm]{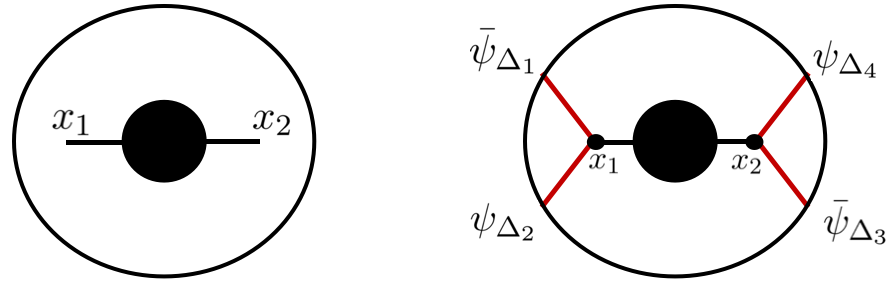}
\caption{\textbf{Left:} The black blob represents a general bulk two-point corelator $F(x_1,x_2)$, with points $x_1$ and $x_2$.  \textbf{Right:} Attaching fermionic bulk-to-boundary propagators (red lines) to the 2-point blob gives a CFT 4-point function. This class of diagrams has 2 bulk-to-boundary propagators attached to $x_1$ and $x_2$, and 2 attached to $x_2$.}
\label{fig:figd1}
\end{figure}

In \cite{Carmi:2019ocp} we considered scalar Witten diagrams, and derived identities which enabled us to compute various families of loop diagrams in AdS. In the current section we consider Witten diagrams containing external fermions. We show that a class of fermionic diagrams is proportional to their scalar companion diagrams.

\subsection{$\langle \bar {\psi} \psi \bar {\psi}  \psi  \rangle$}

Consider a (scalar) 2-point bulk correlator $F(x_1,x_2)$, which is represented by the black blob in Fig.~\ref{fig:figd1}-Left. We consider the correlator of 4 external fermions $\psi$ with scaling dimensions $(\D_1$,  $\D_2$, $\D_3$, $\D_4)$, attached to the 2-point blob $F(x_1,x_2)$, see Fig.~\ref{fig:figd1}-Right. This 4-point function is:
\begin{align}
\langle \bar{\psi}_{\D_1}(P_1,S_1) \psi_{\D_2}(P_2,S_2) \bar {\psi}_{\D_3}(P_3,S_3)  \psi_{\D_4}(P_4,S_4) \rangle
\end{align}
where $P_i$ are points on the boundary of AdS and $S_i$ is a spinor polarization variable. For further details see e.g. section~6 of \cite{Carmi:2018qzm}.
The fermionic bulk-to-boundary propagators $K^F_{\D}$ are given by:
\begin{align}
K^F_{\D} \equiv K^F_{\D} (x,\bar S_b,P,S_\pa) & = \sqrt{ \mm{C}^{F}_{\D} }\frac{  \bar S_b \Pi_- S_\pa  }{(-2x \cdot P)^{\D+\frac{1}{2}}}~,
\\
\bar{K}^F_{\D} \equiv \bar{K}^F_{\D} (x,  S_b,P, \bar S_\pa) & = \sqrt{ \mm{C}^{F}_{\D} } \frac{   \bar S_\pa   \Pi_-   S_b }{(-2x \cdot P)^{\D+\frac{1}{2}}}~,
\\
\text{where}~~~\mm{C}^{F}_{\D}  & \equiv \frac{1}{\pi^{d/2}}  \frac{\G(\D+\frac{1}{2})}{\Gamma(\D+\frac{1-d}{2})}~,
\end{align}
where $x$ is a bulk point, $P$ a boundary point, and $\D$ is the scaling dimension of the fermion. Let us define:
\begin{equation}
S_{12|34} =  ( \bar S_{2\pa} \Pi_- S_{1\pa} ) ( \bar S_{4\pa} \Pi_- S_{3\pa})~,~~ S_{14|32} = (\bar{S}_{4\pa} \Pi_- S_{1\pa} )(  \bar S_{2\pa} \Pi_- S_{3\pa})~.
\end{equation}
where $\Pi_{\pm}$ are chiral projectors in embedding space. The diagram of Fig.~\ref{fig:figd1}-Right is given by:
\begin{align}
S_{12|34}\, \widehat{g}_4^F & = \int d^{d+1} x_1 d^{d+1}x_2  F(x_1,x_2)
  (\pa_{S_{1b}} \pa_{\bar{S}_{2b}}) \bar{K}^F_{\Delta_1}(x_1,  S_{1b},P_1, \bar S_{1\pa}) K^F_{\Delta_2}(x_1,\bar S_{2b},P_2,S_{2\pa})
\nn
& \times(\pa_{S_{3b}} \pa_{\bar{S}_{4b}})  \bar{K}^F_{\Delta_3}(x_2,  S_{3b},P_3, \bar S_{3\pa})K^F_{\Delta_4}(x_2,\bar S_{4b},P_4,S_{4\pa})~,\label{eq:1ovNfer}
\end{align} 
where the integrals $ \int d^{d+1} x_1 d^{d+1}x_2$ are over the $AdS_{d+1}$ space. Now we use the following identity\footnote{See \cite{Kawano:1999au}, and also Eq.~6.19 of \cite{Carmi:2018qzm}.}:
\begin{align}
\label{eq:pldsd}
 & (\pa_{S_{1b}} \pa_{\bar{S}_{2b}})  \bar{K}^F_{\Delta_1}(x_1,  S_{1b},P_1, \bar S_{1\pa}) K^F_{\Delta_2}(x_1,\bar S_{2b},P_2,S_{2\pa}) \nonumber \\ & =\sqrt{(2\Delta_1+1-d)(2\Delta_2+1-d) }(\bar S_{1\pa} \Pi_- S_{2\pa}) K_{\Delta_1+\frac 12}(x_1,P_1) K_{\Delta_2+\frac 12}(x_1,P_2)~,
\end{align}
where $K_{\D}^F$  are fermionic bulk-to-boundary fermionic propagators, and $K_\Delta$ are scalar bulk-to-boundary propagators. This identity enables to write Eq.~\ref{eq:1ovNfer} in terms of scalar propagators:
\begin{align}
\label{eq:pfkso}
S_{12|34}\, \widehat{g}_4^F & = S_{12|34} \Big( \prod_{i=1}^4 \sqrt{(2\Delta_i+1-d)} \Big)
  \int d^{d+1} x_1 d^{d+1} x_2  F(x_1,x_2)
\nn
&\times K_{\delta_1}(P_1,x_1)  K_{\delta_2}(P_2,x_1)  K_{\delta_3}(P_3,x_2)  K_{\delta_4}(P_4,x_2)
\end{align}
where we defined $\d_i=\D_i+\frac{1}{2}$. The second line contains a product of 4 \underline{scalar} bulk-to-boundary propagators with dimension $\d_i$. Thus we showed that the fermionic diagram is equal, up to a factor, to the scalar diagram with shifted scaling dimensions $\d_i=\D_i+\frac{1}{2}$.

\subsubsection*{Example}

Consider the contact tree-level diagram of 4-fermions with scaling dimensions $\D_i$.  Using Eq.~\ref{eq:pfkso} we see that this diagram is proportional to contact diagram of four scalar with scaling dimensions $\d_i= \D_i+\frac{1}{2}$. It is well known that the scalar contact diagram is computed via special functions called $\bar D$ functions, Thus:
\begin{align}
\widehat{g}_4^F (z,\bar z) \sim  \bar D_{\D_1+\frac{1}{2}, \ldots , \D_4+\frac{1}{2}}
\end{align}

\subsection{$\langle \bar {\psi} \psi \phi \phi \rangle$}

\begin{figure}[t]
\centering
\includegraphics[clip,height=3.5cm]{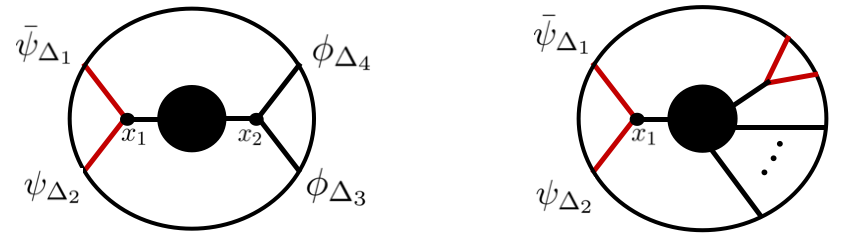}
\caption{\textbf{Left:} Like Fig.~\ref{fig:figd1}-right, only here the $x_2$ vertex has scalar propagators attached to it. \textbf{Right:} A completely general diagram in AdS, which has a vertex $x_1$ to which are attached two fermionic bulk-to-boundary propagators.}\label{fig:figd2}
\end{figure}

Consider the 4-point correlator with 2 external scalars and 2 external fermions, connected to a general 2-point correlator, as in Fig.~\ref{fig:figd2}-Left. We have:
\begin{align}
\label{eq:plff4}
&\int d^{d+1}  x^{d+1} _1 dx_2  \ F(x_1,x_2)
  \nn
&\times (\pa_{S_{1b}} \pa_{\bar{S}_{2b}}) \bar{K}^F_{\Delta_1}(x_1,  S_{1b},P_1, \bar S_{1\pa}) K^F_{\Delta_2}(x_1,\bar S_{2b},P_2,S_{2\pa}) K_{\Delta_3}(P_3,x_2)  K_{\Delta_4}(P_4,x_2)
 \end{align} 
where $F(x_1,x_2)$ is an arbitrary bulk-to-bulk 2-point function. Using the identity of Eq.~\ref{eq:pldsd} in Eq.~\ref{eq:plff4}, gives:
\begin{align}
&  \sqrt{(2\Delta_1+1-d)(2\Delta_2+1-d) }(\bar S_{1\pa} \Pi_- S_{2\pa})
  \nn
 &\times \int d^{d+1}  x_1 d^{d+1}  x_2  F(x_1,x_2) K_{\Delta_1+\frac 12}(x_1,P_1) K_{\Delta_2+\frac 12}(x_1,P_2) K_{\Delta_3}(P_3,x_2)  K_{\Delta_4}(P_4,x_2)
 \end{align} 
which is a 4-point correlator with 4 external scalars.

\subsection{More general fermionic correlators}

Consider a Witten diagram with 2 external fermions attached to a single vertex, as in Fig.~\ref{fig:figd2}-Right. The rest of the diagram is completely arbitrary. Using the identity of Eq.~\ref{eq:pldsd} gives:
\begin{align} 
  \int d^{d+1}  x_1  
  (\pa_{S_{1b}} \pa_{\bar{S}_{2b}}) \bar{K}^F_{\Delta_1}(x_1,  S_{1b},P_1, \bar S_{1\pa}) K^F_{\Delta_2}(x_1,\bar S_{2b},P_2,S_{2\pa})
\times (\ldots)
 \end{align} 
where $(\ldots)$ represents the rest of the diagram, which is arbitrary. Using the identity of Eq.~\ref{eq:pldsd} gives
\begin{align}
&  \sqrt{(2\Delta_1+1-d)(2\Delta_2+1-d) }(\bar S_{1\pa} \Pi_- S_{2\pa})
  \nn
& \times  \int d^{d+1}  x_1  K_{\Delta_1+\frac 12}(x_1,P_1) K_{\Delta_2+\frac 12}(x_1,P_2) \times (\ldots)
 \end{align} 
Therefore a vertex containing 2 external fermions can be replaced with a vertex containing 2 external scalars, and the rest of the diagram stays the same.
As a result, many of the identities derived in \cite{Carmi:2019ocp} for external scalars, can be applied to diagrams with external fermions.

\subsection{Spectral representation}

In this section we recall the spectral representation of the 4-point correlator of 4 fermions, attached to a 2-point function in the bulk as in Fig.~\ref{fig:figd1}-Right. We denote this correlator by $\widehat{g}_4^F(z,\bar z)$, and define the stripped fermionic correlator $g_4(z, \bar z)$ as follows:
\begin{align}
\label{eq:pfkso2}
g_4 (z,\bar z)\equiv \frac{\widehat{g}_4^F}{A^F_{\d_i}}
\end{align}
where we defined the prefactor\footnote{See Eqs.~C.10-C.11 of \cite{Carmi:2019ocp}.}:
\begin{align} 
\label{eq:prefactor}
A^F_{\d_i} \equiv c \frac{\frac{\Big( \frac{P^2_{14}}{P^2_{24}}\Big)^{\frac{\d_2-\d_1}{2}}\Big( \frac{P^2_{14}}{P^2_{13}}\Big)^{\frac{\d_3-\d_4}{2}}}{(P_{12})^{\frac{\delta_1+\d_2}{2}}(P_{34})^{\frac{\delta_3+\d_4}{2}}} \Big( \prod_{i=1}^4 \sqrt{(2\delta_i-d)} \Big) }{\sqrt{\Gamma_{\delta_1}\Gamma_{\delta_2}\Gamma_{\delta_3}\Gamma_{\delta_4}\Gamma_{\delta_1-\frac{d}{2}+1}\Gamma_{\delta_2-\frac{d}{2}+1}\Gamma_{\delta_3-\frac{d}{2}+1}\Gamma_{\delta_4-\frac{d}{2}+1}} }
\end{align}
where $c$ is a numerical factor which will be unimportant for us. 

In Eq.~\ref{eq:pfkso} the fermionic diagram was directly related to that of 4 scalars. Now one can write the spectral representation of this (see Eq.~6.20 of \cite{Carmi:2018qzm}.). The integrand in Eq.~\ref{eq:pfkso} contains the bulk 2-point function $F(x_1,x_2)$, which has a spectral representation:
\begin{align} 
\label{eq:dfdpp}
F(x_1,x_2)=  \int_{-\infty}^\infty d\n  \tilde{F}(\n)  \Omega _{\nu}(x_1,x_2) \,,
\end{align}
where the AdS harmonic function $\Omega _{\nu}(x_1,x_2)$ is a linear combination of a bulk-to-bulk propagator and it's shadow partner:
 \begin{align} 
\label{eq:ksjnf2}
&\O_\n (x_1,x_2) =  \frac{i\n}{2\pi} \Big( G_{\frac{d}{2}+i\n}(x_1,x_2) -G_{\frac{d}{2}-i\n}(x_1,x_2) \Big)
\end{align} 
One can now derive the spectral representation of the 4-point function. The result is:
\begin{align} 
\label{eq:fermg4}
&g_4(z, \bar z)= \int_{-\infty}^\infty d\nu  \tilde{F}(\n) 
\times \frac{  \G_{\frac{\delta_1+\d_2}{2} +\frac{i \nu-\frac{d}{2}}{2}} \G_{\frac{\delta_1+\d_2}{2} -\frac{i \nu+\frac{d}{2}}{2}} \G_{\frac{\delta_3+\d_4}{2} +\frac{i \nu-\frac{d}{2}}{2}} \G_{\frac{\delta_3+\d_4}{2} -\frac{i \nu+\frac{d}{2}}{2}}  }{\Gamma_{i\nu} \Gamma_{\frac{d}{2}+i\n} }
\nn
& \Big( \G_{\frac{\d_2-\delta_1}{2}+\frac{i \nu+\frac{d}{2}}{2}} \G_{\frac{\delta_1-\d_2}{2}+\frac{i \nu+\frac{d}{2}}{2}} \G_{\frac{\delta_3-\d_4}{2}+\frac{i \nu+\frac{d}{2}}{2}} \G_{\frac{\d_4-\delta_3 }{2}+\frac{i \nu+\frac{d}{2}}{2}} \Big) \mathcal{K}^{\d_i}_{\frac{d}{2}+i\nu} (z,\bar z)
\end{align}

where $ \mathcal{K}^{\d_i}_{\frac{d}{2}+i\nu} (z,\bar z)$ is the scalar conformal block with scaling dimensions $\d_i=\D_i+\frac{1}{2}$, and $\tilde{F}(\n)$ is the spectral representation of the bulk 2-point function $F(x_1,x_2)$, Eq.~\ref{eq:dfdpp}.
If in Eq.~\ref{eq:fermg4} one chooses to close the contour in the $\n$-plane and pick up the residues of the poles, one would get the conformal block expansion of the 4-point correlator.

\section{The Gross-Neveu model on $AdS_3$}
\label{sec:grossneveu}

\begin{figure}[t]
\centering
\includegraphics[clip,height=3.2cm]{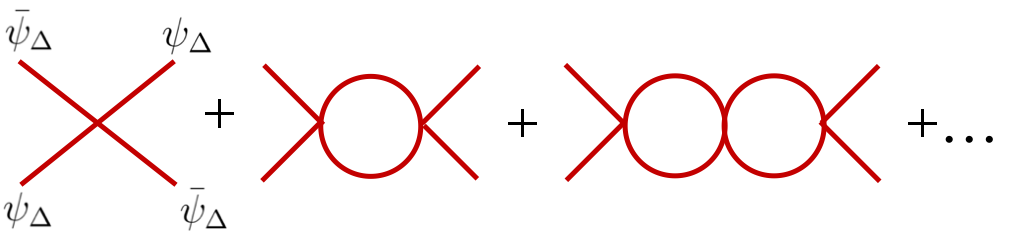}
\caption{The Gross-Neveu model on AdS. The 4-point correlator at large-$N$ is given by an infinite sum of bubble diagrams. The red lines are fermionic propagators. The boundary of AdS is not shown in the figure.} \label{fig:figd3} 
\end{figure}

Consider the Gross-Neveu \cite{Gross:1974jv} model containing $N$ spin-$1/2$ Dirac fermions $\psi^i$ , with $i=1, \ldots, N$, see also \cite{ZinnJustin:1991yn,Rosenstein:1990nm}. In flat-space and large-$N$ this is a solvable model, exhibiting chiral symmetry breaking and asymptotic freedom. The lagrangian is:
\begin{align}
\mm{L} = \bar{\psi}^i \gamma \cdot \nabla \psi^i +   m\bar{\psi}^i \psi^i +\frac{g}{2N} (\bar{\psi}^i \psi^i)^2
\end{align}
We denote the scaling dimension of the $\psi$'s as $\D$.
In \cite{Carmi:2018qzm} we considered the Gross-Neveu model on $AdS_{d+1}$, and computed the finite coupling 4-point correlator in the spectral representation (This is the same as Eq.~\ref{eq:fermg4} with $\D_i =\D$.): 
\begin{align}
\label{eq:123} 
g_4(z, \bar z)=  \int  d\n  \tilde{F}(\n) \frac{\Gamma_{\Delta+\frac{1}{2}-\frac{d+2i\nu}{4}}^2\Gamma_{\Delta+\frac{1}{2}-\frac{d-2i\nu}{4}}^2\Gamma_{\frac{d+2i\nu}{4}}^4}{\Gamma_{i\nu}\Gamma_{\frac{d}{2}+i\nu}}\mathcal{K}^\D_{\frac{d}{2}+i\nu}(z,\bar{z}) \,,
\end{align}
where the spectral function $\tilde{F}(\n)$ is a resummation of 1-loop bubble diagrams, as in Fig.~\ref{fig:figd3}. In this section we compute the $\n$ integral in Eq.~\ref{eq:123}  and thus derive explicitly an expression for the 4-point correlator $g_4(z, \bar z)$ in position-space.

The spectral function $\tilde{F}(\n)$ was computed \cite{Carmi:2018qzm} explicitly in $d=1,2$. Let us consider the case of $d=2$, i.e $AdS_3$, in which the resummation of bubble diagrams gives:
\begin{align}
 \label{eq:1235}
 \tilde{F}(\n)
 = \frac{1}{\frac{-i(4(\D-1)^2+\n^2)}{8\pi \n} \Big( \psi(\D+\frac{i\n}{2})- \psi(\D-\frac{i\n}{2}) \Big)+\frac{2\D-1}{8\pi(\D-1)}}
\end{align}
where the digamma function is $\psi(x) =d\log \G(x)/dx$, not to be confused with the notation for the fermion $\psi$.  \cite{Carmi:2018qzm} showed evidence for a bulk conformal point when the external scaling dimension is $\D=3/2$. Plugging $\D=3/2$ in Eq.~\ref{eq:1235}, the spectral function simplifies:
\begin{align}
  \label{eq:1236}
 \tilde{F}(\n)
 = \frac{8\n}{(1+\n^2)\tanh(\frac{\pi \n}{2})}= \frac{8(h-1)}{h(h-2)\cot(\frac{\pi h}{2})}
\end{align}
where we defined $h\equiv \frac{d}{2}+i\n$. Now we use an identity for the conformal block (see Eq.~3.5 of \cite{Carmi:2019ocp}.):
\begin{align}
2D_{z,\bar z}\mathcal{K}_{\frac{d}{2}+i\n}(z,\bar{z}) =- (\n^2+\frac{d^2}{4}) \mathcal{K}_{\frac{d}{2}+i\n}(z,\bar{z}) 
\end{align}
where the operator $D_{z,\bar z} \equiv z^2(1-z)\pa_z^2-z^2\pa_z+\bar{z}^2(1-\bar{z})\pa_{\bar{z}}^2-\bar{z}^2\pa_{\bar z}$ for $d=2$.
The scalar conformal block in $d=2$ can be written in terms of LegendreQ functions:
\begin{align}
\label{eq:confblock}
\mathcal{K}_{\b}(z,\bar{z}) = k_\b(z) k_\b(\bar z) = 4 \frac{\G^2(\b)}{\G^4(\frac{\b}{2})} Q_{\frac{\b}{2}-1} (\hat z)  Q_{\frac{\b}{2}-1} (\hat{\bar z})
\end{align} 
where $k_\b(z)\equiv z^{\frac{\b}{2}} {}_2 F_1 (\frac{\b}{2}+a,\frac{\b}{2}+b,\b,z)$, $a\equiv \frac{\D_{21}}{2}=0$ and $b\equiv \frac{\D_{34}}{2}=0$, and we defined:
\begin{align}
\hat z \equiv \frac{2}{z}-1 \ \ \ \ \ , \ \ \ \ \ \ \hat{\bar{z}} \equiv \frac{2}{\bar z}-1
\end{align}

Combining Eqs.~\ref{eq:123}-\ref{eq:confblock}, gives:
\begin{align}
g_4(z, \bar z)=   D_{z,\bar z} \int  d\n \frac{ -8\pi^2 \n^2}{\sinh (\pi \n)}  Q_{\frac{i\n-1}{2}} (\hat z)  Q_{\frac{i\n-1}{2}} (\hat{\bar z})
\end{align}
Now we close the contour in the $\n$-plane and use the residue theorem. The poles come from the denominator $\sinh (\pi \n)$, and are at $i\n+1=3+n$, with $n=0,1, \ldots$. This gives:
\begin{align} 
g_4(z,\bar z )= 16\pi^3D_{z,\bar z }\bigg[ \sum_{n=0}^\infty   (2n+1)^2 Q_{n } (\hat z)  Q_{n } (\hat{\bar z}) -  \sum_{n=0}^\infty   (2n+2)^2 Q_{n+\frac{1}{2}} (\hat z)  Q_{n+\frac{1}{2}} (\hat{\bar z})  \bigg]
\end{align}
The square brackets above were computed in equation 4.18 of \cite{Carmi:2019ocp}, in terms of the tree-level contact Witten diagram $g^{contact}_4(z,\bar z )$ of scalars with scaling dimensions $\D_i=(1,1,\frac{3}{2},\frac{3}{2})$. Thus,
\begin{align}
\boxed{
g_4(z,\bar z )= \frac{\pi}{8} D_{z,\bar z } \ g^{contact}_4(z,\bar z ) }
\end{align}
The finite coupling large-$N$ 4-point correlator of the conformal Gross-Neveu model on $AdS_3$, which is given by an infinite sum of fermionic bubble diagrams in Fig.~\ref{fig:figd3}, was computed in terms of a scalar tree-level contact diagram $g^{contact}_4(z,\bar z )$!
In \cite{Carmi:2019ocp}, the 4-point correlator in the conformal $O(N)$ model on $AdS_3$ was computed in terms of the same contact diagram $g^{contact}_4(z,\bar z )$. It would be interesting to understand whether or not this is a coincidence.

\section{Ladder Diagrams in AdS}
\label{sec:ladder}

In this section we derive the spectral representation for $AdS$ ladder diagrams, with an arbitrary number of rungs. We consider only scalars in this section. The ladder diagrams are written as a gluing of tree-level exchange 4-point diagrams. This is possible due to two properties: 1. The spectral/split representation of the bulk-to-bulk propagator. 2. The 6J-symbol expansion of a tree-level exchange diagram in cross channel conformal blocks. See \cite{Liu:2018jhs,Meltzer:2019nbs}

\subsection{Warmup: Tree-level comb Witten diagrams}
\label{sec:tree}

\begin{figure}[t]
\centering
\includegraphics[clip,height=4.0cm]{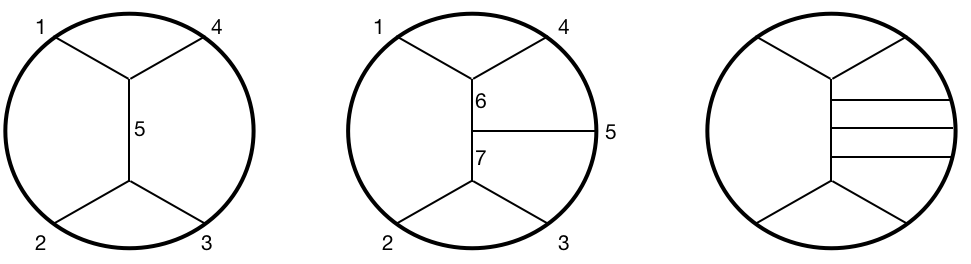}
\caption{\textbf{Left:} 4-point tree-level exchange diagram. \textbf{Middle:} 5-point tree-level exchange diagram. \textbf{Right:} $N$-point tree-level exchange diagram diagram in the comb channel. }\label{fig:figd4} 
\end{figure}

We consider here the comb channel tree-level diagrams in AdS, Fig~\ref{fig:figd4}.
The 4-point exchange Witten diagram, Fig~\ref{fig:figd4}-Left, is given by:
\begin{align} 
\label{eq:fnnf564}
&\mm{A}^{3214}_{5 , exch}=\int d^{d+1} x_1 d^{d+1} x_2  G_{\D_{5}}(x_1,x_2) 
\nn
&\times K_{\D_1}(P_1,x_1)K_{\D_4}(P_4,x_1)K_{\D_3}(P_3,x_2)K_{\D_2}(P_2,x_2)  
\end{align} 
where $G_{\D_{5}}(x_1,x_2)$ is a bulk-to-bulk propagator. The spectral representation gives:
\begin{align}
\label{eq:dn9j9}
\mm{A}^{3214}_{5 , exch}  = \int_{-\infty}^{\infty} d \m_5 \frac{\pi \m_5^2}{ \m_5^2 +(\D_5-\frac{d}{2})^2} b_{32\m_5} b_{- \m_514} \Psi^{3214}_{\m_5}
\end{align}
where $b_{123}$ is defined in Eq.~\ref{eq:nvmmd3}, and the 4-point conformal partial wave is:
\begin{align}
\Psi_{5}^{1234} = \int d^{d+1}x_5 \langle \mm{O}_1 \mm{O}_2  \mm{O}_5 \rangle \langle \tilde{\mm{O}}_5 \mm{O}_3 \mm{O}_4 \rangle 
\end{align}
In a similar manner, the tree-level 5-point comb diagram, Fig~\ref{fig:figd4}-Middle,  gives:
\begin{align}
\mm{A}^{(5)}_{tree} = \int_{-\infty}^\infty \frac{d\m_6 d \m_7  \ \m_6^2 \m_7^2}{[\m^2_{6}+(\D_{6}-\frac{d}{2})^2][\m_7^2 +(\D_7-\frac{d}{2})^2]  }  b_{12 \m_6}b_{-\m_{6}, 5, -\m_{7}} b_{\m_{7}34}\Psi_{\m_6, \m_7}^{12345}
\end{align}
where the 5-point conformal partial wave is:
\begin{align}
\Psi_{67}^{12345} = \int d^{d+1}x_6d^{d+1}x_7 \langle \mm{O}_1 \mm{O}_2  \mm{O}_6 \rangle \langle \tilde{\mm{O}}_6 \mm{O}_5 \tilde{\mm{O}}_7 \rangle \langle \mm{O}_7 \mm{O}_3  \mm{O}_4 \rangle
\end{align}
One can extend these results to $N$-point tree-level comb diagrams, Fig~\ref{fig:figd4}-Right. The result is:
\begin{align}
\mm{A}^{(N)}_{tree}   =  \int_{-\infty}^\infty  \bigg( \prod_{horiz.} \frac{d \m_i \m_i^2   }{ \m_i^2+(\D_i-\frac{d}{2})^2 }  \bigg) \Big( \prod_{vertices} b_{ijk} \Big)    \Psi^{(N-point)}
\end{align}
where the $N$-point conformal partial wave is:
\begin{align}
&\Psi_{N+1 \ldots 2N-3}^{(N-point)} = 
\nn
& \int  d^{d+1}x_{N+1}\cdots d^{d+1}x_{2N-3} 
\langle \mm{O}_1 \mm{O}_2  \mm{O}_{N+1} \rangle \langle \tilde{\mm{O}}_{N+1} \mm{O}_3 \tilde{\mm{O}}_{N+2} \rangle \cdots  \langle \mm{O}_{2N-3} \mm{O}_{N-1}  \mm{O}_N \rangle
\end{align}


\subsection{4-point ladder diagrams}

\begin{figure}[t]
\centering
\includegraphics[clip,height=4.4cm]{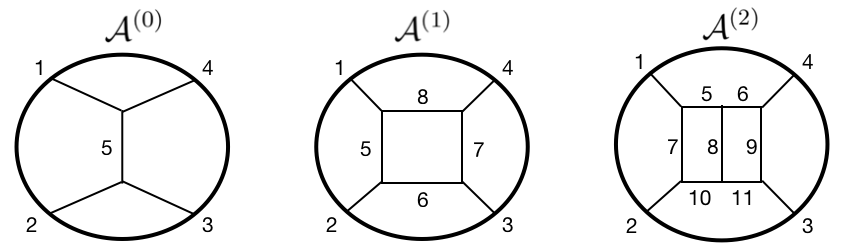}
\caption{4-point ladder diagrams. \textbf{Left:} The tree-level exchange diagram is the 0-loop ladder. \textbf{Middle:} The box diagram is the 1-loop ladder diagram.  \textbf{Right:} The 2-loop ladder diagram.}\label{fig:figd5} 
\end{figure}

The exchange diagram (Fig.~\ref{fig:figd5}-Left) expanded in the direct channel is (Eq.~\ref{eq:dn9j9}):
\begin{align}
\mm{A}^{3214}_{5 , exch} =  \int_{-\infty}^{\infty} d \m_5 \frac{\pi \m_5^2}{ \m_5^2 +(\D_5-\frac{d}{2})^2} b_{32\m_5} b_{- \m_514} \Psi^{3214}_{\m_5}
\end{align}
The same diagram can be expanded  \cite{Liu:2018jhs,Meltzer:2019nbs} in the cross channel conformal partial waves $\Psi^{1234}_{\m,J}$:
\begin{align}
\mm{A}^{3214}_{5 , exch} = b_{32 5} b_{\tilde{ 5}14}\sum^\infty_{J =0} \int \frac{d\m}{2\pi i}(d-2\D_{5})   K^{14}_{\tilde 5}    \left(\begin{array}{c} \mm{O}_1,\mm{O}_2,\mm{O}_5\ \  \\   \mm{O}_3,\mm{O}_4,\mm{O}_{\m,J } \end{array}\right) \frac{1}{n_{\mm O_{\m,J}}} \Psi^{1234}_{\m,J}
\end{align}
Let us define the factor appearing in the integrand above:
\begin{align}
\mm{J}^{1234}_{\mm{O}_5 ,\mm{O}_{\m,J}} \equiv (d-2\D_{5}) b_{32 5} b_{\tilde{ 5}14} K^{14}_{\tilde 5}\left(\begin{array}{c} \mm{O}_1,\mm{O}_2,\mm{O}_5\ \ \  \\   \mm{O}_3,\mm{O}_4,\mm{O}_{\m ,J }  \end{array}\right)  \frac{1}{n_{\mm O_{\m,J}}}
\end{align}
Therefore we can write:
\begin{align}
\label{eq:lkd3}
\mm{A}^{(0)}=\mm{A}^{3214}_{5 , exch} =\sum^\infty_{J =0} \int \frac{d\m}{2\pi i}\ \mm{J}^{1234}_{\mm{O}_5 ,\mm{O}_{\m,J}}  \ \Psi^{1234}_{ \m,J}
\end{align}
This exchange diagram is the lowest order (zero loop) 4-point ladder diagram, Fig.~\ref{fig:figd5}-Left. In order to simplify the notation, we suppress writing the exchanged $\mm{O}_5 $ operator, namely we write: 
\begin{align}
\mm{J}^{1234}_{ \m,J} \equiv \mm{J}^{1234}_{\mm{O}_5,\mm O_{\m,J}} 
\end{align}

The 1-loop ladder is the box diagram, Fig~\ref{fig:figd5}-Middle. The box diagram can be computed as a gluing of two exchange diagrams:
\begin{align}
\mm{A}^{(1)} =\mm{A}_{Box} = \mm{A}^{621\tilde{\underline{8}}}_{5 , exch}  \otimes \mm{A}^{3\tilde{\underline{6}}\underline{8}4}_{7, exch} 
\end{align}
Similarily, the 2-loop ladder (Fig~\ref{fig:figd5}-Right) is a gluing of three exchange diagrams:
\begin{align}
\mm{A}^{(2)} =  \mm{A}^{\underline{10} , 2, 1, \underline{5}}_{7 , exch}  \otimes \mm{A}^{\underline{11}, \tilde{\underline{10}} ,\tilde{\underline{5}},\underline{6}   }_{8, exch}  \otimes  \mm{A}^{3, \tilde{\underline{11}} ,\tilde{\underline{6}}, 4   }_{8, exch}
\end{align}
The $N$-loop ladder diagram is a gluing of $N+1$ exchange diagrams. Schematically:
\begin{align}
\mm{A}^{(N)} = \mm{A}_{1 , exch}  \otimes  \mm{A}_{2 , exch} \ \otimes \ \cdots \  \otimes \ \mm{A}_{N+1 , exch}
\end{align}
To derive an explicit expression for the $N$-loop ladder diagram, it is useful to write it as a conformal partial wave decomposition: 
\begin{align}
\label{eq:pldnnd6}
\mm{A}^{(N)} = \ \sum_{J =0}^\infty \int_{-\infty}^\infty \frac{d\m}{2\pi i} \   C^{(N)}_{\m,J} \Psi^{1234}_{\m,J}
\end{align} 
where $C^{(N)}_{\m,J}$ is the OPE function at $N$-loops, and $\Psi^{1234}_{\m,J}$ is the 4-point conformal partial wave. For the tree-level exchange diagram we have from Eq.~\ref{eq:lkd3}:
\begin{align}
\label{eq:pldnnd666}
C^{(0)}_{\m,J}  =  \mm{J}^{\zz{1},\zz{2},\zz{3},\zz{4}}_{ \m,J}    
\end{align}
The 1-loop ladder OPE function is:
\begin{align}
\label{eq:flkkd4}
C^{(1)}_{\m,J}  =  B_{ \m,J } \int_{-\infty}^\infty  \bigg( \prod_{i=6,8} \frac{d \n_i \n_i^2   }{ \n_i^2+(\D_i-\frac{d}{2})^2 }  \bigg)   \mm{J}^{\zz{1},\zz{2},6,8}_{ \m,J}   \mm{J}^{\tilde 8,\tilde{6},\zz{3},\zz{4}}_{ \m,J}
\end{align}
We derive this in Appendix \ref{sec:AA2}. For a somewhat different derivation, see \cite{Meltzer:2019nbs}. $B_{ \m,J }$ is a conformal bubble factor. The external operators in $\mm{J}$ are highlighted in red. Notice that $i$ runs over the horizontal bulk-to-bulk propagators, see Fig.~\ref{fig:figd5}. 

In a similar fashion, one can derive the OPE function of the 2-loop ladder Fig.~\ref{fig:figd5}-Right:
\begin{align}
C^{(2)}_{\m,J}  = (B_{ \m,J })^2 \int_{-\infty}^\infty  \bigg( \prod_{i=5,6,10,11} \frac{d \n_i \n_i^2   }{ \n_i^2+(\D_i-\frac{d}{2})^2 }  \bigg)   \mm{J}^{\zz{1},\zz{2},10,5}_{ \m,J} \mm{J}^{\tilde 5,\tilde{10},11,6}_{ \m,J} \mm{J}^{\tilde 6,\tilde{11},\zz{3},\zz{4}}_{ \m,J}
\end{align}
See also \cite{Meltzer:2019nbs}.
We see that there is a product of $\mm J$'s, with their indices integrated over. Let us define the product:
\begin{align}
\label{eq:plmd3}
 \prod_{vertical}    \mm{J}_{ \m,J }=  \mm{J}^{\zz{1},\zz{2},A_1,B_1}_{ \m,J} \mm{J}^{\tilde{B}_1, \tilde{A}_1,A_2,B_2}_{ \m,J} \mm{J}^{\tilde{B}_2, \tilde{A}_2,A_3,B_3}_{ \m,J}\ \ \cdots \ \mm{J}^{\tilde{B}_{\hat N-1}, \tilde{A}_{\hat N-1}, A_{\hat N},B_{\hat N}}_{\ \m,J}  \mm{J}^{\tilde{B}_{\hat N}, \tilde{A}_{\hat N},\zz{3},\zz{4}}_{ \m,J}
\end{align}
The $N$-loop ladder diagram OPE function is given by:
\begin{align}
\label{eq:pol90}
\boxed{
C^{(N-loop)}_{\m,J} =\big( B_{ \m,J }\big)^N
 \int_{-\infty}^\infty  \bigg( \prod_{horiz.} \frac{d \n_i \n_i^2   }{ \n_i^2+(\D_i-\frac{d}{2})^2 }  \bigg)    \prod_{vertical}    \mm{J}_{ \m,J}   }
\end{align}
Notice that all of the spectral integrals over the vertical propagators have been effectively performed, and we are left only with spectral integrals over the horizontal propagators.
There is a 6j-symbol factor $\mm{J}_{ \m,J} $ for each vertical bulk-to-bulk propagator. There is an integral $d\n_i$ for each horizontal bulk-to-bulk propagators.

\subsection{2 and 3-point ladder diagrams}

\begin{figure}[t]
\centering
\includegraphics[clip,height=4.2cm]{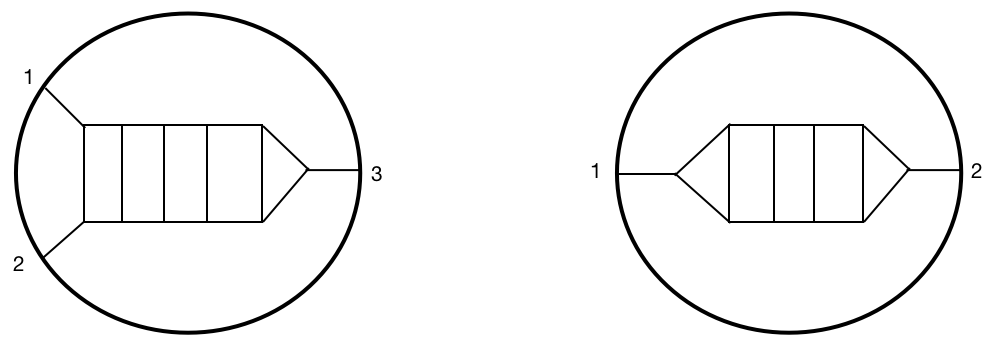}
\caption{\textbf{Left:} An example of a 3-point ladder diagram. \textbf{Right:} An example of a 2-point ladder diagram. }\label{fig:figd6} 
\end{figure}

In the previous subsection we looked at 4-point ladder diagrams in $AdS$. In this subsection we consider ladder 3-point and 2-point functions, Fig.~\ref{fig:figd6}. Thus the gluings for the 3-point ladder are:
\begin{align}
\mm{A}^{(N)}_{(3-point)} = \mm{A}_{(3-point)} \otimes  \mm{A}_{1 , exch} \ \otimes \ \cdots \  \otimes \ \mm{A}_{N , exch}
\end{align}
and similarly for the 2-point functions:
\begin{align}
\mm{A}^{(N)}_{(2-point)} = \mm{A}_{(3-point)} \otimes  \mm{A}_{1 , exch} \ \otimes \ \cdots \  \otimes \ \mm{A}_{N-1 , exch} \  \otimes \ \mm{A}_{(3-point)}
\end{align}
A computation similar to the 4-point case gives:
\begin{align} 
&\mm{A}^{(N)}_{(3-point)} =  C^{(N)}_{(3-point)} \  \langle \mm{O}(P_1) \mm{O}(P_2) \mm{O}(P_3) \rangle
\end{align} 
and
\begin{align} 
&\mm{A}^{(N)}_{(2-point)} =  C^{(N)}_{(2-point)} \  \langle \mm{O}(P_1) \mm{O}(P_2)   \rangle
\end{align} 
Compared to the 4-point ladders of Eq.~\ref{eq:pldnnd6}, the 3-point ladders have fixed space-time dependence given by $\langle O(P_1)O(P_2) O(P_3 )  \rangle$. Similarly the 2-point ladders have 2-point structure $\langle \mm{O}(P_1) \mm{O}(P_2)   \rangle$. The factors $C^{(N)}$above depend on the scaling dimensions (and not on the space-time coordinates), and are given by:
\begin{align}
\boxed{
C^{(N-loop)}_{(3-point)} =
 \int_{-\infty}^\infty  \bigg( \prod_{horiz.} \frac{d \n_i \n_i^2   }{ \n_i^2+(\D_i-\frac{d}{2})^2 }  \bigg) \big( B_{ \m,J }\big)^N   \prod_{vertical}    \mm{J}_{ \m,J }  \ \Big|_{J=0,\m =\D_3-\frac{d}{2}} }
\end{align}
and
\begin{align}
\label{eq:pol901}
\boxed{
C^{(N-loop)}_{(2-point)} =
 \int_{-\infty}^\infty  \bigg( \prod_{horiz.} \frac{d \n_i \n_i^2   }{ \n_i^2+(\D_i-\frac{d}{2})^2 }  \bigg) \big( B_{ \m,J  }\big)^N   \prod_{vertical}    \mm{J}_{\m,J }  \ \Big|_{J=0,\m =\D_1-\frac{d}{2}} }
\end{align}

\subsection{Mixed ladders/bubbles}

One can also compute diagrams  that contain both ladders and bubbles. As an example, consider the diagram in Fig.~\ref{fig:figd7}-Left. It's OPE function is:
\begin{align}
&C_{\m,J}  = (B_{ \m,J })^2\frac{1}{\m^2+(\D_8-\frac{d}{2})^2}\frac{1}{\m^2+(\D_{12}-\frac{d}{2})^2} \widetilde{B}^{(\D_{13},\D_{14})}_\m
\nn
&\times \int_{-\infty}^\infty  \bigg( \prod_{i=5,6,10,11} \frac{d \n_i \n_i^2   }{ \n_i^2+(\D_i-\frac{d}{2})^2 }  \bigg)   \mm{J}^{\zz{1},\zz{2},10,5}_{ \m,J} \mm{J}^{\tilde 5,\tilde{10},6,11}_{ \m,J} b_{\tilde{6},\tilde{11},\m}b_{\tilde \m, \zz 3,\zz 4 }
\end{align}
where $\widetilde{B}^{(\D_{13},\D_{14})}_\m$ is defined as the spectral representation of the 1-loop bubble:
\begin{align}
G_{\D_{13}} (x_1,x_2)G_{\D_{14}} (x_1,x_2) = \int_{-\infty}^\infty d\m \ \widetilde{B}^{(\D_{13},\D_{14})}_\m \O_\m  (x_1,x_2)
\end{align}
where $\O_\m  (x_1,x_2)$ is the AdS harmonic function.

\subsection{$\phi^4$  ladders}

\begin{figure}[t]
\centering
\includegraphics[clip,height=4.6cm]{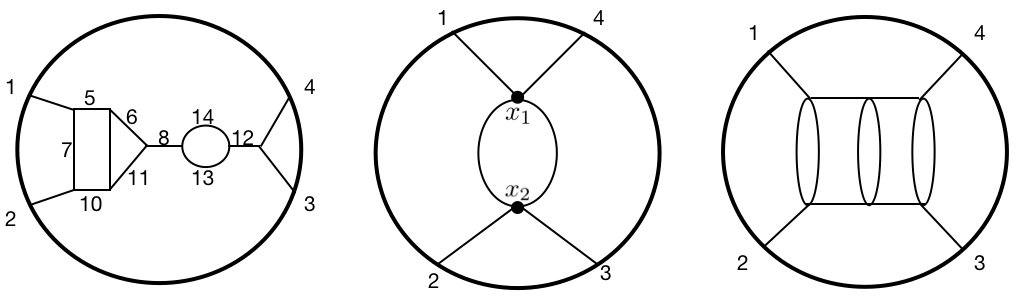}
\caption{\textbf{Left:} An example of a mixed ladder-bubble diagram in $\phi^3$ theory on AdS. \textbf{Middle:} The 4-point 1-loop bubble diagram in $\phi^4$ theory. \textbf{Right:} An example of a $\phi^4$ ladder diagram.}\label{fig:figd7} 
\end{figure}

Let us consider ladder diagrams in $\phi^4$ theory on AdS. First consider the 4-point bubble diagram, Fig~\ref{fig:figd7}-Middle:
\begin{align}
\label{eq:fjj44f}
\mm{A}_{bubble}= \int d^{d+1} x_1 d^{d+1}x_2 \ K_{\D_1}(P_1,x_1)K_{\D_2}(P_2,x_2)K_{\D_3}(P_3,x_2)K_{\D_4}(P_4,x_1) 
G^2_{\D}(x_1,x_2)  
\end{align} 
The bulk propagator squared $G^2_{\D}(x_1,x_2)$ can be expanded as a sum of bulk propagators \cite{Fitzpatrick:2010zm,Fitzpatrick:2011hu}:
\begin{align}
G^2_{\D}(x_1,x_2) = \sum_{n=0}^\infty a_{\D,\D}(n)G_{2\D+2n}(x_1,x_2)  
\end{align}
where the coefficients are:
\begin{align}
\label{eq:fff1232}
a_{\D_1,\D_2}(n) =\frac{ (\frac{d}{2})_n}{2\pi^{\frac{d}{2}}\Gamma_{n+1}} \frac{(\D_1+\D_2+2n)_{1-\frac{d}{2}}(\D_1+\D_2+n-d+1)_n}{ (\D_1+n)_{1-\frac{d}{2}} (\D_2+n)_{1-\frac{d}{2}} (\D_1+\D_2+n-\frac{d}{2})_n}
\end{align}
Thus Eq.~\ref{eq:fjj44f} becomes:
\begin{align}
&\mm{A}_{bubble}= \sum_{n=0}^\infty a_{\D,\D}(n)  
\nn
& \times \int d^{d+1} x_1 d^{d+1}x_2 \ K_{\D_1}(P_1,x_1)K_{\D_2}(P_2,x_2)K_{\D_3}(P_3,x_2)K_{\D_4}(P_4,x_1) 
G_{2\D+2n}(x_1,x_2) 
\end{align} 
The second line above is just a tree-level exchange diagram, see Eq.~\ref{eq:fnnf564}. 
\begin{align}
&\mm{A}_{bubble}= \sum_{n=0}^\infty a_{\D,\D}(n)  \mm{A}^{1234}_{\mm{O}' , exch} \Big|_{\D_{\mm O'} = 2\D +2n} 
\end{align} 
Combining this with Eqs.~\ref{eq:pldnnd6} and \ref{eq:pldnnd666}, gives the OPE function:
\begin{align}
\label{eq:lkjjhg43}
(C^{(0)}_{\m,J})_{bubble} = \sum_{n=0}^\infty a_{\D,\D}(n) C^{(0)}_{\m,J}  \Big|_{\D_{\mm O'} \to 2\D +2n}=  \sum_{n=0}^\infty a_{\D,\D}(n) \mm{J}^{\zz{1},\zz{2},\zz{3},\zz{4}}_{\mm{O}',\mm{O}_{\m,J}}    \Big|_{\D_{\mm O'} \to 2\D+2n} 
\end{align}
Let us define the sum:
\begin{align}
\mm{S}^{\zz{1},\zz{2},\zz{3},\zz{4}}_{\mm{O}',\mm{O}_{\m,J}}  \equiv \sum_{n=0}^\infty a_{\D,\D}(n) \mm{J}^{\zz{1},\zz{2},\zz{3},\zz{4}}_{\mm{O}',\mm{O}_{\m,J}}    \Big|_{\D_{\mm O'} \to 2\D+2n}
\end{align}
Now can simply write Eq,~\ref{eq:lkjjhg43} as:
\begin{align}
(C^{(0)}_{\m,J})_{bubble} =\mm{S}^{\zz{1},\zz{2},\zz{3},\zz{4}}_{\mm{O}',\mm{O}_{\m,J}}
\end{align}

One can easily extend this to higher loop $\phi^4$ ladders, as in Fig.~\ref{fig:figd7}-Right. All one has to do is make the replacement $\mm{J} \to \mm{S}$. Thus instead of Eq.~\ref{eq:pol90}, we have the OPE function:
\begin{align}
\boxed{
(C^{(N)}_{\m,J} )_{bubble}=\big( B_{ \m,J }\big)^N
 \int_{-\infty}^\infty  \bigg( \prod_{horiz.} \frac{d \n_i \n_i^2   }{ \n_i^2+(\D_i-\frac{d}{2})^2 }  \bigg)    \prod_{vertical\ bubbles}    \mm{S}_{ \m,J}   }
\end{align}
where the product is defined as:
\begin{align}
 \prod_{vertical\ bubbles}    \mm{S}_{ \m,J }=  \mm{S}^{\zz{1},\zz{2},A_1,B_1}_{ \m,J} \mm{S}^{\tilde{B}_1, \tilde{A}_1,A_2,B_2}_{ \m,J} \mm{S}^{\tilde{B}_2, \tilde{A}_2,A_3,B_3}_{ \m,J}\ \ \cdots \ \mm{S}^{\tilde{B}_{\hat N-1}, \tilde{A}_{\hat N-1}, A_{\hat N},B_{\hat N}}_{\ \m,J}  \mm{S}^{\tilde{B}_{\hat N}, \tilde{A}_{\hat N},\zz{3},\zz{4}}_{ \m,J}
\end{align}

\section{2-point bulk correlators in AdS}
\label{sec:2point6}

\begin{figure}[t]
\centering
\includegraphics[clip,height=4.3cm]{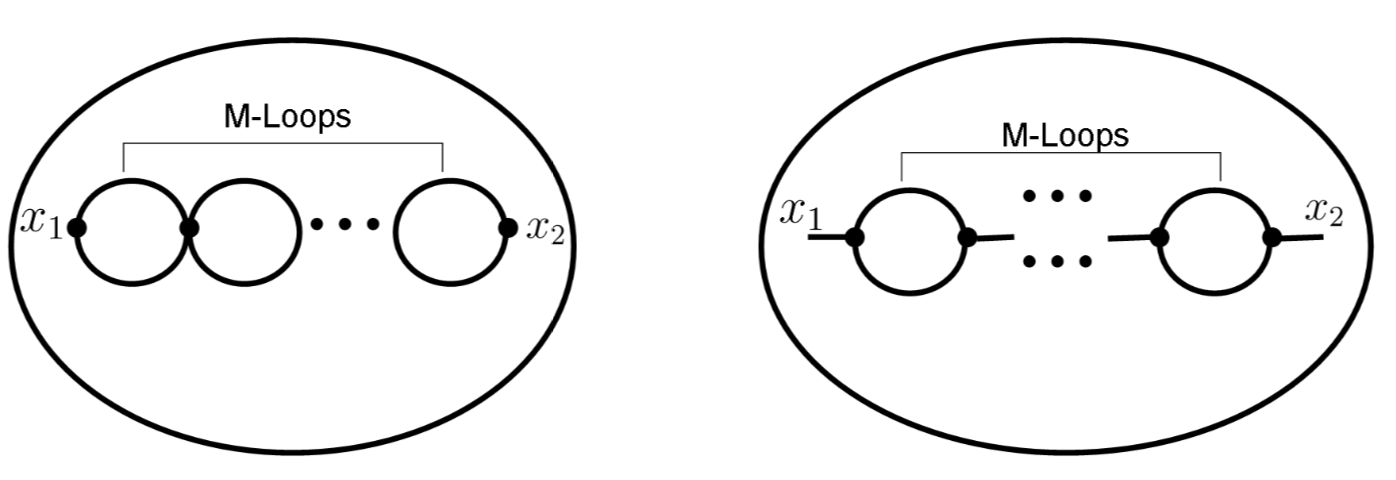}
\caption{The type of bubble diagrams which we consider in section~\ref{sec:2point6}. \textbf{Left:} $\phi^4$ theory: The 2-point bulk-to-bulk correlator consisting of a sequence of $M$-bubbles.  \textbf{Right:} $\phi^3$ theory: The 2-point bulk-to-bulk correlator consisting of a sequence of $M$-bubbles.  }
\label{fig:figd8} 
\end{figure}
Consider 2-point bulk bubble diagrams of scalars in $AdS_3$, with $x_1$ and $x_2$ being the external points in the bulk, such as those in Fig~\ref{fig:figd8}. The $AdS_{d+1}$ bulk-to-bulk scalar propagator in position space is:
\begin{align}
G_\D(x_1,x_2) =  \frac{\G_\D}{2\pi^{\frac{d}{2}} \G_{\D-\frac{d}{2}+1}} \zeta^{-\D} {}_2F_1 (\D,\D-\frac{d-1}{2},2\D-d+1,-4\zeta^{-1})
\end{align}
where $\zeta= \frac{(z_1-z_2)^2+(\vec{x}_1-\vec{x}_2)^2}{z_1z_2}$ is the chordal distance squared between the points $x_1$ and $x_2$. When $d$ is even, the propagator above simplifies. In particular, for $d=2$ ($AdS_3$):
\begin{align}
\label{eq:ksjnf23}
G_\D(x_1,x_2) =  \frac{1}{2\pi} \frac{1}{\sqrt{\zeta(\zeta+4)}}\Big( \frac{2}{\sqrt{\zeta}+\sqrt{\zeta+4}} \Big)^{2\D-2}
=   \frac{1}{2\pi} \frac{1}{\sqrt{\zeta(\zeta+4)}} \eta^{\D-1}
\end{align}
i.e the bulk-to-bulk propagator is a power law in $\D$.
We defined $\eta \equiv \Big( \frac{2}{\sqrt{\zeta}+\sqrt{\zeta+4}} \Big)^2$, which has the range $\eta$ is $0 \leq \eta \leq1$.
Likewise, in $AdS_5$ the bulk-to-bulk propagator becomes a power law in $\D$:
\begin{align}
\label{eq:ksjnf239}
G_\D(x_1,x_2) =  (A+B \D )\omega^\D
\end{align}
where we defined:
\begin{align}
&\o\equiv 4\zeta^{-1}\Big(1+\sqrt{\frac{4+\zeta}{\zeta}}\Big)^{-2}
\nn
&A\equiv \frac{  \Big(1+\sqrt{\frac{4+\zeta}{\zeta}}\Big)^{3 } }{64\pi^{2}  \Big(\frac{4+\zeta}{\zeta} \Big)^{\frac{3}{2}} }  \Big[ -2 \Big(1+\sqrt{\frac{4+\zeta}{\zeta}}\Big)+\frac{4}{\zeta}\Big( -3+\sqrt{\frac{4+\zeta}{\zeta}}\Big) \Big] 
\nn 
&B\equiv \frac{  \Big(1+\sqrt{\frac{4+\zeta}{\zeta}}\Big)^{3 } }{64\pi^{2}  \Big(\frac{4+\zeta}{\zeta} \Big)^{\frac{3}{2}} }  \Big[ 2 \Big(1+\sqrt{\frac{4+\zeta}{\zeta}}\Big)+\frac{8}{\zeta}  \Big] 
\end{align}

The spectral representation of the bulk-to-bulk propagator is:
\begin{align} 
\label{eq:bulkspect49}
G_\D (x_1,x_2) = \int_{-\infty}^\infty \frac{ d\nu}{\n^2+(\D-\frac{d}{2})^2}  \Omega _{\nu}(x_1,x_2) 
\end{align} 

where the AdS harmonic function is:
 \begin{align} 
\label{eq:ksjnf2}
&\O_\n (x_1,x_2) =  \frac{i\n}{2\pi} \Big( G_{\frac{d}{2}+i\n}(x_1,x_2) -G_{\frac{d}{2}-i\n}(x_1,x_2) \Big)
\nn
&= \frac{h-\frac{d}{2}}{2\pi} \Big( G_{h}(x_1,x_2) -G_{d-h}(x_1,x_2) \Big)
\end{align} 
where $h\equiv i\n+\frac{d}{2}$.

\subsection*{$\phi^4$ bulk bubble diagrams in $AdS_5$}
\label{eq:nolegs}
Consider a scalar field in AdS with $\phi^4$ interaction. The  bubble diagrams Fig.~\ref{fig:figd8}-Left  give a contribution to the 2-point function:
\begin{align}
\langle \phi^2(x_1) \phi^2(x_2) \rangle_{bulk}
\end{align}
The spectral representation of a sequence of $M$ bubbles is just the $M$th power of a single bubble.
\begin{align} 
\label{eq:dkja45}
&g_2^{(M)}(x_1,x_2) =  \int_{-\infty}^\infty d\nu ( \tilde{B}(\n) )^M  \Omega _{\nu}(x_1,x_2) = \frac{1}{\pi} \int_{-\infty}^\infty d\nu ( \tilde{B}(\n) )^M  i \n G_{\frac{d}{2}+i\nu}(x_1,x_2) 
\end{align}
where $G_{\frac{d}{2}+i\nu}(x_1,x_2) $ is the bulk-to-bulk propagator, and we used Eq.~\ref{eq:ksjnf2}. We go to position space by closing the $\n$ contour and using the residue theorem. For a pole of order $M$ at $y=y_0$, the residue is:
\begin{align}
Res(F(y))\Big|_{y=y_0} = \frac{1}{\G_M}  \frac{d^{M-1}}{dy^{M-1} } \Big[(y-y_0)^M F(y) \Big] \Big|_{y\to y_0}
\end{align}
Thus Eq.~\ref{eq:dkja45} gives:
\begin{align} 
\label{eq:workwork}
&g_2^{(M)}(x_1,x_2) =   \frac{2}{\G_M}\sum^\infty_{n=0}  \frac{d}{d^{M-1} h} \bigg[ (h-(2n+2\D))^M ( \tilde{B}(\n) )^M  (h-\frac{d}{2}) G_{h}(x_1,x_2) \bigg]_{h\to (2n+2\D)}
\end{align}
where $h\equiv \frac{d}{2}+i\n$. As an example, consider the regularized bubble with $\D=2$ in $AdS_5$, Eq.~\ref{eq:pldnnndd8}:

\begin{align}
\tilde{B}_{reg.}(\n)= \frac{\n^3 \coth(\frac{\pi \n}{2})}{1+\n^2} = \frac{(h-2)^3\cot \frac{\pi h}{2}}{(h-1)(h-3)}  
\end{align}
Plugging this in Eq.~\ref{eq:workwork} gives:
\begin{align} 
\label{eq:workwork2}
&g_2^{(M)}(x_1,x_2) =   \frac{2}{\G_M}\sum^\infty_{n=0}   \frac{d}{d^{M-1} h} \bigg[ \Big( (h-(2n+4)) \frac{(h-2)^3\cot \frac{\pi h}{2}}{(h-1)(h-3)}  \Big)^M    (h-2)  (A+B h )\omega^h  \bigg]_{h\to (2n+4)}
\end{align}

For a given value of $M$ (number of loops), one can compute the sum above in terms of hypergeometric functions\footnote{We will leave it for the interested reader to obtain explicit expressions, which can easily be done with Mathematica. In a similar fashion, one can also compute bubble diagrams in $\phi^3$ theory, as in Fig.~\ref{fig:figd8}-Right. }.

\subsection*{Fermionic bubble diagrams}

Consider fermions in the bulk of $AdS_3$ ($d=2$). The poles of the 1-loop bubble are  (see Eq~.6.26 of \cite{Carmi:2018qzm}):
\begin{align}
 \tilde{B}^{(d=2)}_F(\nu) \overset{i\nu\, \sim  \,2\Delta+2n}{\sim}-\frac{1}{i\nu-(2\Delta+2n)} \times \frac{(n+1)(2\D+n-1)}{(2\D+2n)}
 \end{align}

Therefore for a chain of fermionic bubbles:
\begin{align} 
&g^{(M)}(x_1,x_2)=  \int \frac{d\nu}{2\pi}  ( \tilde{B}_{F}(\n) )^M  \Omega _{\nu}(x_1,x_2) \,,
\nn
& =  \sum_{n=0}  \frac{d}{d^{M-1} h} \bigg[ (h-(2n+2\D+1))^M ( \tilde{B}_F(\n) )^M  (h-1) G_{h}(x_1,x_2) \bigg]_{h\to (2n+2\D)}
\nn
& =  \frac{1}{2\pi} \frac{1}{\sqrt{\zeta(\zeta+4)}}  \sum_{n=0}  \frac{d}{d^{M-1} h} \bigg[ (h-(2n+2\D+1))^M ( \tilde{B}_F(\n) )^M  (h-1) \eta^{h-1}  \bigg]_{h\to (2n+2\D+1)}
\end{align}

For $M=1$ computing the sum gives:
\begin{align} 
g^{(M=1)}(x_1,x_2)  =  \frac{1}{2\pi} \frac{1}{\sqrt{\zeta(\zeta+4)}}  \frac{\eta^{2\D}(1-2\D +\eta^2(2\D-3))}{(\eta^2-1)^3}
\end{align}

One can continue and compute higher loop bubble diagrams.

\subsection*{Sunset bubbles}

\begin{figure}[t]
\centering
\includegraphics[clip,height=3.6cm]{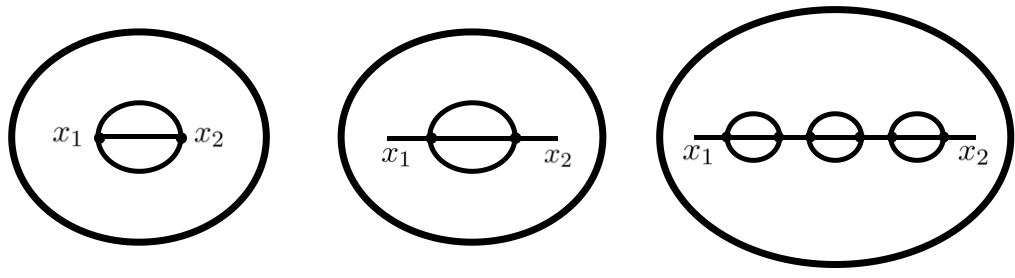}
\caption{Sunset diagrams. \textbf{Left:} The sunset bubble. \textbf{Middle:} $\phi^4$ theory bulk 2-point function $\langle \phi \phi \rangle_{bulk}$ at 2-loops. \textbf{Right:} Chain of sunset bubble diagrams in $\phi^4$. }
\label{fig:figd9} 
\end{figure}

Consider the 2-loop bubble between 2-points in the bulk $x_1$ and $x_2$, otherwise known as the sunset diagram, see Fig.~\ref{fig:figd9}-Left. In position space this diagram is simply equal to the propagator cubed: $G^3_{\D}(x_1,x_2)$. In Eq.~55 of \cite{Fitzpatrick:2011hu}, this was written in terms of an infinite sum over bulk propagators: 
\begin{align}
\label{eq:nc4f5}
G^3_{\D}(x_1,x_2) = \sum_{n=0}^\infty N_{3,\D}(n)G_{3\D+2n}(x_1,x_2)  
\end{align}
where
\begin{align}
N_{3,\D}(n)=\sum_{m=0}^n a_{\D,\D}(m) a_{\D,2\D+2m}(n-m)  
\end{align}
and the coefficients $a_{\D_1,\D_2}$ are defined in Eq.~\ref{eq:fff1232}.
Plugging Eq.~\ref{eq:bulkspect49} in Eq.~\ref{eq:nc4f5} gives:
\begin{align}
&G^3_{\D}(x_1,x_2) =   \sum_{n=0}^\infty N_{3,\D}(n) G_{3\D+2n}(x_1,x_2)
\nn
&=  \sum_{n=0}^\infty N_{3,\D}(n) \int_{-\infty}^\infty \frac{ d\nu}{\n^2+(3\D+2n-\frac{d}{2})^2}  \Omega _{\nu}(x_1,x_2) 
\nn
&= \int_{-\infty}^\infty  d\nu   \Omega _{\nu}(x_1,x_2)  \sum_{n=0}^\infty\frac{ N_{3,\D}(n)}{\n^2+(3\D+2n-\frac{d}{2})^2}
\nn
&\equiv \int_{-\infty}^\infty  d\nu   \Omega _{\nu}(x_1,x_2)  B_{sun}(\n)
\end{align}
where $B_{sun}(\n)$ is the spectral representation of $G^3_{\D}(x_1,x_2) $. Therefore the spectral representation of the sunset bubble is:
\begin{align}
\label{eq:sun2}
 B_{sun}(\n) =\sum_{n=0}^\infty\frac{ N_{3,\D}(n)}{\n^2+(3\D+2n-\frac{d}{2})^2}
\end{align}
Now we focus on $AdS_3$ ($d=2$) where we have a simplification:
\begin{align}
a^{(d=2)}_{\D_1,\D_2}(n) =\frac{1}{2\pi}
\end{align}
thus
\begin{align}
N^{(d=2)}_{3,\D}(n)=\sum_{m=0}^n a_{\D,\D}(m) a_{\D,2\D+2m}(n-m) = \frac{1}{4\pi^2} \sum_{m=0}^n 1=\frac{n+1}{4\pi^2}
\end{align}
Therefore Eq.~\ref{eq:sun2} gives:
\begin{align}
\label{eq:mdkjjkd}
B^{(d=2)}_{sun}(\n) =   \frac{1}{4\pi^2}\sum_{n=0}^\infty  \frac{n+1}{\n^2 + (3\D+2n-1)^2 }
\end{align}
This sum is logarithmically divergent and therefore should be regularized. Clearly the poles of the sunset bubble are:
\begin{align}
\label{eq:dino90}
 \tilde{B}_{sun}^{(d=2)}(\nu) \overset{1+i\nu\, \sim  \,3\Delta+2n}{\sim}-\frac{1}{1+i\nu-(3\Delta+2n)} \times \frac{n+1}{(3\D+2n-1)}
\end{align}

\subsubsection*{Consistency check}
 
Let's compute the sum in Eq.~\ref{eq:nc4f5}, and see that the result is consistent:
\begin{align}
\label{eq:nc4f}
G^3_{\D}(x_1,x_2) = \sum_{n=0}^\infty N_{3,\D}(n)G_{3\D+2n}(x_1,x_2)  
\nn
 =\frac{1}{(2\pi )^3\sqrt{\zeta(\zeta+4)}} \sum^\infty_{n=0}  (n+1)\eta^{3\D+2n-1} 
 \nn
=  \Big( \frac{1}{2\pi} \frac{1}{\sqrt{\zeta(\zeta+4)}} \eta^{\D-1} \Big)^3 = G^3_{\D}(x_1,x_2)
\end{align}
 where in the first equality we used Eqs.~\ref{eq:ksjnf23}. The sum in the second line is just a geometric sum. We get the propagator cubed, and everything is consistent.

\subsubsection*{Chain of sunsets}

Consider the 2-point bulk correlator composed of a chain of $M$ sunset bubbles such as that in Fig.~\ref{fig:figd9}-Right:
\begin{align} 
g^{(M)}(x_1,x_2)=  \int \frac{d\nu}{2\pi}    \frac{( \tilde{B}_{sun}(\n) )^M}{\Big( \n^2+(\d-\frac{d}{2})^2\Big)^{M+1}}\Omega _{\nu}(x_1,x_2) \,,
\end{align}
For simplicity let's consider $AdS_3$ and  $M=1$ (Fig.~\ref{fig:figd9}-Middle):
\begin{align} 
\label{eq:mdkjjkd2}
&g^{(1)}(x_1,x_2)=   \int \frac{d\nu}{2\pi}  \frac{\tilde{B}_{sun}(\n)}{\Big((h+\d-2)( h-\d)\Big)^{2}}   \Omega _{\nu}(x_1,x_2)
\nn
& =   \frac{1}{2\pi \sqrt{\zeta(\zeta+4)}} \sum^\infty_{n=0}  \frac{(n+1)\eta^{3\D+2n-1}}{\Big((2n+3\D+\d-2)( 2n+3\D-\d )\Big)^{2}}
\nn
&+  \frac{1}{2\pi} \frac{1}{\sqrt{\zeta(\zeta+4)}}  \frac{d}{dh}\Big[  \frac{(h-1)}{(h+\d-2)^2} \tilde{B}^{sun}(\n)\eta^{h-1}\Big] \bigg|_{h =\d }
\nn
\end{align}
where the last line above comes from the double pole $\sim \frac{1}{(h-\d)^2}$. The last line can be computed by plugging $\tilde{B}^{sun}(\n)$ from Eq.~\ref{eq:mdkjjkd}. Note that one should first regularize the sum in Eq.~\ref{eq:mdkjjkd}. 

The sum in the second line of Eq.~\ref{eq:mdkjjkd2} can be computed as:
\begin{align}
&\frac{1}{2\pi \sqrt{\zeta(\zeta+4)}} \sum^\infty_{n=0}  \frac{(n+1)\eta^{3\D+2n-1}}{\Big((2n+3\D+\d-2)( 2n+3\D-\d )\Big)^{2}}
 \nn
& =  \frac{1}{2\pi \sqrt{\zeta(\zeta+4)}} 
\frac{\eta^{3\D-1}}{32  (\d-1)^3} \Big[ 6(\D-1) \Phi( \eta^2, 1,\frac{3\D-\d}{2})-6(\D-1)\Phi( \eta^2, 1,\frac{3\D+\d-2}{2} )
\nn
&+ (\d-1)\Big( (2+\d-3\D) \Phi( \eta^2, 2,\frac{3\D-\d}{2})+(4-\d-3\D  ) \Phi( \eta^2, 2,\frac{3\D+\d-2}{2} ) \Big) \Big]
\end{align}
Where $\Phi$ is the Lerch transcendent function defined as:
\begin{align}
\Phi(y,s,\a) \equiv \sum_{n=0}^\infty \frac{y^n}{(n+\a)^s}
\end{align}

\section*{Acknowledgements}

I am grateful to Lorenzo Di Pietro, Shota Komatsu, Eric Perlmutter, David Meltzer, Alic Sivaramakrishnan for useful discussions.


\appendix

\section{Ladder diagram at 1-loop}
\label{sec:AA2}

\begin{figure}[t]
\centering
\includegraphics[clip,height=4.5cm]{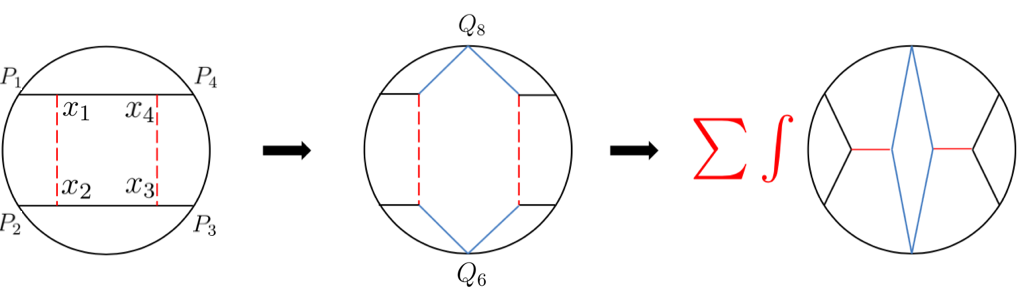}
\caption{Schematic explanation of the computation of the one-loop box diagram. First, we rewrite the two horizontal bulk-to-bulk propagators using the split representation. The result is given by a convolution of two tree-level exchange diagrams. Second, we write exchange diagrams in the cross-channel conformal block expansion by using the $6j$ symbol.}\label{fig:figd10} 
\end{figure}

In this section we show the details of the calculation of the spectral representation of the 1-loop ladder diagram.
The 4-point 1-loop ladder diagram is given by (see Fig.~\ref{fig:figd5}-Middle):
\begin{align}
\label{eq:fnndd}
\mm{A}^{(1)}= \int d^{d+1} x_1 \cdots d^{d+1}x_4 \ K_{\D_1}(P_1,x_1)K_{\D_2}(P_2,x_2)K_{\D_3}(P_3,x_3)K_{\D_4}(P_4,x_4) 
\nn
G_{\D_5}(x_1,x_2) G_{\D_7}(x_3,x_4)\  G_{\D_8} (x_1, x_4)\   G_{\D_6} (x_2, x_3)~.
\end{align} 
We use the spectral representation of two of the bulk-to-bulk propagators
\begin{align}
G_{\D_8} (x_1, x_4) & =  \int_{-\infty}^\infty d\n_8 \frac{1}{\n_8^2+(\D_8-\frac{d}{2})^2} \O_{\n_8}(x_1, x_4)~, \\
G_{\D_6} (x_2, x_3) & =  \int_{-\infty}^\infty d\n_6 \frac{1}{\n_6^2+(\D_6-\frac{d}{2})^2} \O_{\n_6}(x_2, x_3)~,
\end{align} 
and the split representation for AdS harmonic functions:
\begin{align}
\O_{\n_8}(x_1, x_4) &= \frac{\n_8^2    }{\pi} \int d^dQ_8 K_{\frac{d}{2}-i\n_8 } (Q_8,x_1)K_{\frac{d}{2}+i\n_8 } (Q_8,x_4)~,\\
\O_{\n_6}(x_2, x_3) & = \frac{\n_6^2  }{\pi} \int d^dQ_6 K_{\frac{d}{2}+i\n_6} (Q_6,x_2)K_{\frac{d}{2}-i\n_2} (Q_6,x_3)~,
\end{align} 
where $Q_i$ are boundary points. This procedure is schematically illustrated in Fig.~\ref{fig:figd10}. Therefore Eq.~\ref{eq:fnndd}  becomes:
\begin{align}
&\mm{A}^{(1)}=  \frac{1}{\pi^2}  \prod_{i=6,8} \int_{-\infty}^\infty d\n_i  \int d^dQ_i  \frac{ \n_i^2   }{ \n_i^2+(\D_i-\frac{d}{2})^2 } 
\nn
&\int d^{d+1}x_1 d^{d+1}x_2 \  K_{\D_1}(P_1,x_1)K_{\D_2}(P_2,x_2)    K_{\frac{d}{2}-i\n_8 } (Q_8,x_1)K_{\frac{d}{2}+i\n_6 } (Q_6,x_2)  G_{\D_5}(x_1,x_2)
\nn
&\int d^{d+1}x_3 d^{d+1}x_4 \ K_{\D_3}(P_3,x_3)K_{\D_4}(P_4,x_4) K_{\frac{d}{2}+i\n_8 } (Q_8,x_4)K_{\frac{d}{2}- i\n_6} (Q_6,x_3)   G_{\D_7}(x_3,x_4)
\end{align}
The bottom 2 lines are tree-level exchange diagrams, thus:
\begin{align}
\mm{A}^{(1)}= \frac{1}{\pi^2}  \prod_{i=6,8} \int_{-\infty}^\infty d\n_i  \int d^dQ_i  \frac{ \n_i^2   }{ \n_i^2+(\D_i-\frac{d}{2})^2 }\ \  \mm{A}^{621\tilde{8}}_{5 , exch} \otimes  \mm{A}^{3\tilde{6}84 }_{7 , exch}
\end{align}
The exchange diagrams have a conformal partial wave expansion in the crossed channel (Eq.~\ref{eq:lkd3}):
\begin{align}
\mm{A}^{621\tilde{8}}_{\mm{O}_5 , exch} =\sum^\infty_{J_5 =0} \int \frac{d\m_5}{2\pi i}\ \mm{J}^{\zz{12}6\tilde{8}}_{\mm{O}_5 ,O_{\m_5,J_5}}  \ \Psi^{126\tilde{8}}_{ \m_5,J_5}
\nn
\mm{A}^{3\tilde{6}84}_{\mm{O}_7 , exch} =\sum^\infty_{J_7 =0} \int \frac{d\m_7}{2\pi i}\ \mm{J}^{8\tilde{6}\zz{34}}_{\mm{O}_7 ,O_{\m_7,J_7}}  \ \Psi^{8\tilde{6}34}_{ \m_7,J_7}
\end{align}
Thus,
\begin{align}
\label{eq:fkdds}
&\mm{A}^{(1)} =\frac{1}{\pi^2} \int_{-\infty}^\infty  \bigg( \prod_{i=6,8}   d\n_i  \frac{ \n_i^2   }{ \n_i^2+(\D_i-\frac{d}{2})^2 } \bigg) \sum^\infty_{J_5 =0}\sum^\infty_{J_7 =0} \int \frac{d\m_5}{2\pi i}  \int \frac{d\m_7}{2\pi i}  \mm{J}^{\zz{12}6\tilde{8}}_{\mm{O}_5 ,O_{\m_5,J_5}}       \mm{J}^{8\tilde{6}\zz{34}}_{\mm{O}_7 ,O_{\m_7,J_7}}  
\nn 
&\times \int d^dQ_8 d^dQ_6  \Psi^{126\tilde{8}}_{ \m_5,J_5}    \Psi^{8\tilde{6}34}_{ \m_7,J_7}
\end{align}
The conformal partial waves have a shadow representation:
\begin{align}
\label{eq:pol266}
&\Psi^{126\tilde{8}}_{ \m_5,J_5}    =  \int d^dP_0 \langle O(P_1)O(P_2) O_{\m_5, J_5}(P_0 ) \rangle  \langle  \tilde{O}_{\m_5, J_5}(P_0)\tilde{O}(Q_8)O(Q_6) \rangle
\nn
&\Psi^{8\tilde{6}34}_{ \m_7,J_7}  =  \!\!\int d^dP_0' \langle O(Q_8) \tilde{O}(Q_6) O_{\m_7, J_7}(P_0 ') \rangle  \langle  \tilde{O}_{\m_7, J_7}(P_0 ')O(P_3)O(P_4) \rangle
\end{align} 
Plugging this in Eq.~\ref{eq:fkdds}, we can perform the $Q_8$, $Q_8$ integrals as follows\footnote{This is a CFT bubble integral. The bubble factor $B_{\m_5,J_5}$ is a simple known function, defined e.g in appendix A of \cite{Meltzer:2019nbs}}:
\begin{align}
\label{eq:pol4}
&\int d^dQ_8 d^dQ_6 \langle  \tilde{O}_{\m_5, J_5}(P_0)\tilde{O}(Q_8)O(Q_6) \rangle \langle  O(Q_8)\tilde{O}(Q_6) O_{\m_7, J_7}(P_0 ') \rangle
\nn
&= B_{\m_5,J_5} \d_{P_0,P_0'} \d_{\m_5,\m_7} \d_{J_5,J_7}~.
\end{align} 
Thus:
\begin{align}
\int d^d Q_8 d^dQ_8  \Psi^{126\tilde{8}}_{ \m_5,J_5}    \Psi^{8\tilde{6}34}_{ \m_7,J_7}
=B_{\m_5,J_5}    \Psi^{1234}_{\m_5, J_5} \d_{\m_5,\m_7} \d_{J_5,J_7}
\end{align} 
and we get:
\begin{align}
 \frac{1}{\pi^2}    \sum^\infty_{J_5 =0} \int \frac{d\m_5}{2\pi i}  \bigg[ B_{\m_5,J_5}   \int_{-\infty}^\infty \Big( \prod_{i=6,8} d\n_i  \frac{ \n_i^2   }{ \n_i^2+(\D_i-\frac{d}{2})^2 }  \Big)   \mm{J}^{\zz{12}6\tilde{8}}_{\mm{O}_5 ,O_{\m_5,J_5}}       \mm{J}^{8\tilde{6}\zz{34}}_{\mm{O}_7 ,O_{\m_5,J_5}} \bigg] \Psi^{1234}_{\m_5, J_5}
\end{align}
The external operators in $\mm{J}$ are highlighted in red. The factor in the square brackets above gives the OPE function:
\begin{align}
C^{(1)}_{\m_5,J_5}  = B_{\m_5,J_5}   \int_{-\infty}^\infty \Big( \prod_{i=6,8} d\n_i  \frac{ \n_i^2   }{ \n_i^2+(\D_i-\frac{d}{2})^2 }  \Big)   \mm{J}^{\zz{12}6\tilde{8}}_{\mm{O}_5 ,O_{\m_5,J_5}}       \mm{J}^{8\tilde{6}\zz{34}}_{\mm{O}_7 ,O_{\m_5,J_5}}
\end{align}
Thus we derived Eq.~\ref{eq:flkkd4}.

\subsection*{Conformal factors:}

Throughout this discussion of ladder diagrams we mainly use the notation of \cite{Meltzer:2019nbs}, see appendix A therein.
The conformal 2-point function is:
\begin{align}
\langle \mm{O}_{\D,J} (P_1, z_1) \mm{O}_{\D,J} (P_2, z_2) \rangle = \mm{C}_{\D,J} \frac{(z_1\cdot I(P_{12}) \cdot z_2)^J}{P_{12}^{2\D_{\mm{O}}}}
\end{align}
with the definitions
\begin{align}
\mm{C}_\D \equiv  \frac{\G_\D}{2\pi^{\frac{d}{2}} \G_{\D+1-\frac{d}{2}}}
\nn
I_{\m \n} (x) = \d_{\m \n} - 2\frac{x_\m x_\n}{x^2}
\end{align}
The AdS tree-level 3-point correlator:
\begin{align}
\int_{AdS} d^{d+1}x K_{\D_1} (P_1,x)K_{\D_2} (P_2,x)K_{\D_3,J_3} (P_3,x) =b_{123} \langle \mm{O}_1(P_1) \mm{O}_2(P_2) \mm{O}_3(P_3)\rangle
\end{align}
where we defined:
\begin{align}
\label{eq:nvmmd3}
b_{123} \equiv \mm{C}_{\D_1} \mm{C}_{\D_2} \mm{C}_{\D_3,J_3} \frac{\pi^{\frac{d}{2}} \G_{\frac{\D_1+\D_2+\D_3+J_3-d}{2}}  \G_{\frac{-\D_1+\D_2+\D_3+J_3}{2}}  \G_{\frac{\D_1+\D_2-\D_3+J_3}{2}}  \G_{\frac{\D_1-\D_2+\D_3+J_3}{2}}}{2^{1-J_3}\G_{\D_1}\G_{\D_2}\G_{\D_3+J_3}}
\end{align}

\section{An eigenvalue equation for ladder diagrams?}
\label{sec:b2}

Let us consider the 4-point ladder with all scaling dimensions equal to $\D$. Let's define $\n_0\equiv \D-\frac{d}{2}$, and rewrite Eq.~\ref{eq:pol90}:
\begin{align}
\frac{1}{\big( B_{ \m,J }\big)^N} C^{(N-loop)}_{\m,J} =
\big( B_{ \m,J }\big)^N \int_{-\infty}^\infty  \bigg( \prod_{horiz.} \frac{d \n_i \n_i^2   }{ \n_i^2+(\D-\frac{d}{2})^2 }  \bigg)     \prod_{vertical}    \mm{J}_{ \m,J }  
\end{align}
Recalling Eq.~\ref{eq:plmd3} we can view\footnote{I thank L. Di Pietro and S. Komatsu for initial collaboration on AdS ladder diagrams.} the equation above as a ``matrix product" of $\mm{J}_{ \m,J }  $'s
\begin{align}
\label{eq:bncbnb2}
\frac{1}{\big( B_{ \m,J }\big)^N}C^{(N-loop)}_{\m,J} = \Phi^T  \cdot \widehat{\mm{J}}^{N-1} \cdot  \Phi~,  
\end{align} 
where 
\begin{equation}
\Phi_{(\n_1,\n_2)}   \equiv \mm{J}_{\m ,J }^{-\n_2, -\n_1, \n_0,\n_0}~~~~,~~ ~~~\Phi^{T\,(\n_1,\n_2)}   \equiv \mm{J}_{\m ,J }^{\n_0,\n_0,\n_1,\n_2}~,
\end{equation}
is viewed as a vector and 
\begin{equation}
\widehat{\mm{J}}_{(\n_1,\n_2)}^{\phantom{(\n_1,\n_2)}(\n_3,\n_4)}\equiv\mm{J}_{\mu,J}^{-\n_2, -\n_1, \n_3, \n_4}~,
\end{equation}
as a matrix, with indices given by couples of $\nu$ variables. The matrix multiplication is given by the integrals
\begin{equation}
(\widehat{\mm{J}} \cdot \Phi)_{(\n_1,\n_2)} \equiv \int_{-\infty}^{\infty} \int_{-\infty}^{\infty}  d\n_3 d\n_4 \,  \frac{ \n_3^2   }{ \n_3^2+(\D-\frac{d}{2})^2 } \frac{ \n_4^2   }{ \n_4^2+(\D-\frac{d}{2})^2 } ~~\widehat{\mm{J}}_{(\n_1,\n_2)}^{\phantom{(\n_1,\n_2)}(\n_3,\n_4)} \, \Phi_{(\n_3,\n_4)}~.
\end{equation}
Suppose we can diagonalize the matrix $\widehat{\mm{J}}$, i.e. we find a ``basis" (in some appropriate sense) of eigenvectors $\phi_i$ satisfying
\begin{align}
(\widehat{\mm{J}} \cdot \phi_i)_{(\n_1,\n_2)} & = \lambda_i  ~\phi_{i\,(\n_1,\n_2)} ~, 
\end{align} 
where $\lambda_i$'s are the eigenvalues, and let us also assume that we can normalize the eigenvectors as
\begin{equation}
\phi_i^T \cdot \phi_j = \delta_{ij}~.
\end{equation}
We can then expand the vector $\Phi$ in this basis
\begin{align}
\Phi & = \sum_i \a_i \phi_i~.
\end{align}
In terms of these $\lambda_i$'s and $\a_i$'s, it is then immediate to write the OPE function\footnote{The matrix product is: 
\begin{align}
& \Phi^T  \cdot \widehat{\mm{J}}^{N-1} \cdot  \Phi
=( \sum_i \a_i \phi_i)^T \cdot \widehat{\mm{J}}^{N-1} \cdot ( \sum_j \a_j \phi_j)
\nn
&= \sum_i \sum_j  \a_i \a_j  \l_j^{N-1}  \phi_i^T \cdot \phi_j = \sum_i \sum_j  \a_i \a_j  \l_j^{N-1}  \d_{ij}
 =\sum_i (\a_i)^2 \l_i^{N-1} 
\end{align} 
} of Eq.~\ref{eq:bncbnb2}:
\begin{align}
\frac{1}{\big( B_{ \m,J }\big)^N}C^{(N-loop)}_{\m,J} = \Phi^T  \cdot \widehat{\mm{J}}^{N-1} \cdot  \Phi
 =\sum_i (\a_i)^2 \l_i^{N-1} 
\end{align}

\section{The $O(N)$ model on $AdS_5$}
\label{sec:c2}

In this section we consider the large-$N$ $O(N)$ model on $AdS_5$. The lagrangian of the $O(N)$ model is:
\begin{align}
\mm{L} = \frac{1}{2} (\pa \phi^i)^2 +\frac{m^2}{2}( \phi^i)^2+\frac{\l}{2N} (\phi^i)^4
\end{align}


 In contrast to the $AdS_3$ case (\cite{Carmi:2019ocp,Carmi:2018qzm}), in $AdS_5$ we will need to regularize the 1-loop bubble. The 4-point correlator at large-$N$ is given by a resummation of bubble diagrams\footnote{The resummation of bubble diagrams is as in Fig~\ref{fig:figd3}. In the spectral representation, the resummation gives a geometric sum:
\begin{align}
\l+ \l^2(-2\tilde{B}(\n)) + \l^3(-2\tilde{B}(\n))^2 +\ldots = \frac{1}{\l^{-1}+2\tilde{B}(\n)}
\end{align}
This explains the factor in Eq.~\ref{eq:ldppf}. } (\cite{Carmi:2018qzm,Carmi:2019ocp}):
\begin{align}
\label{eq:ldppf}
g_4(z, \bar z)= \int_{-\infty}^\infty d\n \frac{1}{\l^{-1}+2\tilde{B}(\n)}\frac{\Gamma_{\Delta-\frac{d+2i\nu}{4}}^2\Gamma_{\Delta-\frac{d-2i\nu}{4}}^2\Gamma_{\frac{d+2i\nu}{4}}^4}{ \Gamma_{i\nu}\Gamma_{\frac{d}{2}+i\nu}}\mathcal{K}_{\frac{d}{2}+i\nu}(z,\bar{z}) 
\end{align}
where $\tilde{B}(\n)$ is the 1-loop bubble in the spectral representation. We will focus on the case of very strong coupling, in which $\l\to \infty$.
 $\tilde B(\n)$ has single poles at (see e.g. Eq.~4.24 of \cite{Carmi:2018qzm}):
 \begin{align}
\tilde{B}(\nu)\overset{\frac{d}{2}+i\nu\, \sim  \,2\Delta+2n}{\sim}-\frac{1}{\frac{d}{2}+i\nu-(2\Delta+2n)}\frac{(\frac{d}{2})_n\Gamma_{\Delta+n}\Gamma_{\Delta+n-\frac{d}{2}+\frac{1}{2}}\Gamma_{2\Delta+n-\frac{d}{2}}}{2(4\pi)^{\frac{d}{2}}\Gamma_{n+1}\Gamma_{\Delta+n+\frac{1}{2}}\Gamma_{\Delta+n-\frac{d}{2}+1}\Gamma_{2\Delta-d+n+1}}\,.
 \end{align}
Focusing on $AdS_5$, we plug $d=4$ above and get:
 \begin{align}
 \label{eq:nff0f00}
 \tilde{B}(\nu)\overset{\frac{d}{2}+i\nu\, \sim  \,2\Delta+2n}{\sim}-\frac{1}{\frac{d}{2}+i\nu-(2\Delta+2n)} \times \frac{(n+1)(n+\D-1)(n+2\D-3)}{2(2\D+2n-1)(2\D+2n-3)} 
 \end{align}
To obtain $\tilde{B}(\nu)$, one should sum over all the poles in Eq.~\ref{eq:nff0f00}. Since this sum is divergent, we regularize it by first subtracting the summand with $\n=0$, and then sum over the poles:
\begin{align}
 \tilde{B}_{reg.}(\nu) \equiv \sum_{n=0}^\infty \bigg[ Eq.~(\ref{eq:nff0f00}) - \Big( Eq.~(\ref{eq:nff0f00}) \Big|_{\n=0} \Big) \bigg]
\end{align}
The regularized bubble $\tilde{B}_{reg.}(\nu)$ is finite. Performing the sum above gives:
\begin{align}
 \tilde{B}_{reg.}(\nu)= \frac{(5-2\D)\n^2}{2(1+\n^2)} +\frac{i \n(\n^2+4(\D-2)^2)}{4(1+\n^2)} \Big[ \psi (-1+\D-\frac{i \n}{2}) - \psi (-1+\D+\frac{i \n}{2}) \Big]
\end{align}
Where $\psi (x)$ is the digamma function. For simplicity, let us focus on the case $\D=2$, in which the regularized bubble simplifies further:
\begin{align}
\label{eq:pldnnndd8}
\tilde{B}_{reg.}(\n)= \frac{\pi}{4} \frac{\n^3 \coth(\frac{\pi \n}{2})}{1+\n^2}  
\end{align}
In $d=4$, the conformal block in Eq.~\ref{eq:ldppf} can be written in terms of LegendreQ functions:
\begin{align}
&\mathcal{K}_{\b}(z,\bar{z}) = \frac{z\bar z}{\bar z -z} \Big[ k_{\b-2}(z) k_\b(\bar z) - k_{\b-2}(\bar z) k_\b( z) \Big] 
\nn
&=  \frac{z\bar z}{\bar z -z} 4 \frac{\G(\b)\G(\b-2)}{\G^2(\frac{\b-2}{2})\G^2(\frac{\b}{2})} \Big[ Q_{\frac{\b}{2}-1} (\hat z)  Q_{\frac{\b}{2}-2} (\hat{\bar z})-Q_{\frac{\b}{2}-2} (\hat z)  Q_{\frac{\b}{2}-1} (\hat{\bar z}) \Big]
\end{align} 
Plugging this in Eq.~\ref{eq:ldppf}, one gets:
\begin{align}
\label{eq:nnnnf0f}
g_4(z, \bar z)=  \frac{z\bar z}{\bar z -z}\int d\n  \frac{-\pi \n (1+\n^2)}{\sinh(\pi \n)} \Big[ Q_{\frac{i\n}{2}} (\hat z) Q_{\frac{i\n}{2}-1} (\hat{\bar z}) - Q_{\frac{i\n}{2}}(\hat{\bar z}) Q_{\frac{i\n}{2}-1} (\hat{ z})  \Big]
\end{align}
We rewrite the square brackets above, using the recurrence relation $Q_{\b-1}(x)= (x+\frac{1-x^2}{\b} \pa_x)Q_\b(x)$:
\begin{align}
 Q_{\frac{i\n}{2}} (\hat{z})Q_{\frac{i\n}{2}-1} (\hat{\bar z}) -Q_{\frac{i\n}{2}-1} (\hat{z})Q_{\frac{i\n}{2}} (\hat{\bar z})=\Big[ \frac{\hat{z}^2-1}{\frac{i\n}{2}} \pa_{\hat{z}} - \frac{\hat{\bar{z}}^2-1}{\frac{i\n}{2}} \pa_{\hat{\bar{z}}} +(\hat{\bar{z}} -\hat{z})\Big] Q_{\frac{i\n}{2}} (\hat{z})Q_{\frac{i\n}{2}} (\hat{\bar z}) 
\end{align}
Thus
\begin{align}
\label{eq:fjjff2}
g_4(z, \bar z)=  \frac{-\pi i z\bar z}{\bar z -z} \int d\n  \frac{ i \n(i\n+1)(i\n-1)}{\sinh(\pi \n)} \Big[ \frac{\hat{z}^2-1}{\frac{i\n}{2}} \pa_{\hat{z}} - \frac{\hat{\bar{z}}^2-1}{\frac{i\n}{2}} \pa_{\hat{\bar{z}}} +(\hat{\bar{z}} -\hat{z})\Big] Q_{\frac{i\n}{2}} (\hat{z})Q_{\frac{i\n}{2}} (\hat{\bar z})
\end{align}
Now we use the following identity for Legendre functions:
\begin{align}
\label{eq:pdpderp}
w_z Q_{\frac{\b}{2}} = \frac{\b}{2}(\frac{\b}{2}+1)Q_{\frac{\b}{2}}  
\nn
\text{where} \ \ \ \ w_z \equiv -\frac{d}{d\hat{z}} (1-\hat{z}^2)\frac{d}{d\hat{z}}
\end{align}
Eq.~\ref{eq:fjjff2} becomes:
\begin{align}
\label{eq:fnnnnf2}
g_4(z, \bar z)=  \mm{P}_1  \int d\n  \frac{  (i\n+1) }{\sinh(\pi \n)}   Q_{\frac{i\n}{2}} (\hat{z})Q_{\frac{i\n}{2}} (\hat{\bar z}) +\mm{P}_2 \int d\n  \frac{  (i\n+1)^2 }{\sinh(\pi \n)}   Q_{\frac{i\n}{2}} (\hat{z})Q_{\frac{i\n}{2}} (\hat{\bar z})
\end{align}
Where we have defined the differential operators:
\begin{align}
\mm{P}_1 = -\pi i \frac{z\bar z}{\bar z -z}  (\hat{\bar{z}} -\hat{z}) (4w_z+3)+4\pi i  \frac{z\bar z}{\bar z -z}  \Big[  (\hat{z}^2-1)  \pa_{\hat{z}} - (\hat{\bar{z}}^2-1) \pa_{\hat{\bar{z}}}  \Big] 
\nn
\mm{P}_2 \equiv 3\pi i \frac{z\bar z}{\bar z -z}  (\hat{\bar{z}} -\hat{z})  -2\pi i  \frac{z\bar z}{\bar z -z}  \Big[  (\hat{z}^2-1)  \pa_{\hat{z}} - (\hat{\bar{z}}^2-1) \pa_{\hat{\bar{z}}}  \Big]
\end{align}
Using the residue theorem on Eq.~\ref{eq:fnnnnf2} gives:
\begin{align}
g_4(z, \bar z)=\pi \mm{P}_1\bigg[ \sum_{n=0}^\infty (2n+2)Q_{n+\frac{1}{2}} (\hat{z}) Q_{n+\frac{1}{2}} (\hat{\bar z}) - \sum_{n=0}^\infty (2n+3)Q_{n+1} (\hat{z}) Q_{n+1} (\hat{\bar z}) \bigg]
\nn
+ \pi \mm{P}_2\bigg[ \sum_{n=0}^\infty (2n+2)^2Q_{n+\frac{1}{2}} (\hat{z}) Q_{n+\frac{1}{2}} (\hat{\bar z}) - \sum_{n=0}^\infty (2n+3)^2Q_{n+1} (\hat{z}) Q_{n+1} (\hat{\bar z}) \bigg]
\nn
= \pi \mm{P}_1S_6  + \pi \mm{P}_2S_7 +\pi \mm{P}_2  Q_{0} (\hat{z}) Q_{0} (\hat{\bar z})
\end{align}
where we defined:
\begin{align}
S_6 \equiv \sum_{n=0}^\infty   (2n+2) Q_{n+\frac{1}{2}} (\hat{z})Q_{n+\frac{1}{2}} (\hat{\bar z})  -  \sum_{n=0}^\infty   (2n+3) Q_{n+1} (\hat{z})Q_{n+1} (\hat{\bar z})
\nn
S_7 \equiv \sum_{n=0}^\infty   (2n+2)^2 Q_{n+\frac{1}{2}} (\hat{z})Q_{n+\frac{1}{2}} (\hat{\bar z})  -  \sum_{n=0}^\infty   (2n+1)^2 Q_{n} (\hat{z})Q_{n} (\hat{\bar z})
\end{align}
These sums can be computed analytically. The $S_6$ sum is equal to (see Eq.~5.9 of \cite{Carmi:2019ocp}):
\begin{align}
S_6 =\Big[2Q_{\frac{1}{2}}(\hat{z}) Q_{\frac{1}{2}}(\hat{\bar z})+\frac{\frac{3}{2}}{(\hat{z}-\hat{\bar z})}\Big( Q_{\frac{1}{2}}(\hat{z}) Q_{\frac{3}{2}}(\hat{\bar z})-Q_{\frac{3}{2}}(\hat{z}) Q_{\frac{1}{2}}(\hat{\bar z}) \Big)\Big]
\nn
- \Big[3Q_1(\hat{z}) Q_1(\hat{\bar z})+\frac{2}{(\hat{z}-\hat{\bar z})}\Big( Q_1(\hat{z}) Q_2(\hat{\bar z})-Q_2(\hat{z}) Q_1(\hat{\bar z}) \Big)\Big]
\end{align}
The sum $S_7$ is equal to (see Eq.~4.18 of \cite{Carmi:2019ocp}):
\begin{align}
S_7 =\frac{1}{128\pi^2} g_4^{(contact)} (\hat{z}, \hat{\bar z})
\end{align}
where $g_4^{(contact)} (\hat{z}, \hat{\bar z})$ is defined as the scalar tree-level contact diagram in $AdS_3$ with external scaling dimensions $\D_i= (1,1,\frac{3}{2},\frac{3}{2})$.
 
To summarize, we have managed to compute analytically the (regularized) 4-point correlator of the $O(N)$ model on $AdS_5$, with scaling dimension $\D=2$:
\begin{align}
\label{eq:nnnnf0f1}
\boxed{
g_4(z, \bar z)= \pi \mm{P}_1S_6  + \pi \mm{P}_2S_7 +\pi \mm{P}_2  Q_{0} (\hat{z}) Q_{0} (\hat{\bar z})}
\end{align}
 where all of these functions were explicitly defined and computed above.\\


\section{Scalar 4-point bubble diagrams in $AdS_5$}
\label{sec:5}

\begin{figure}[t]
\centering
\includegraphics[clip,height=4.2cm]{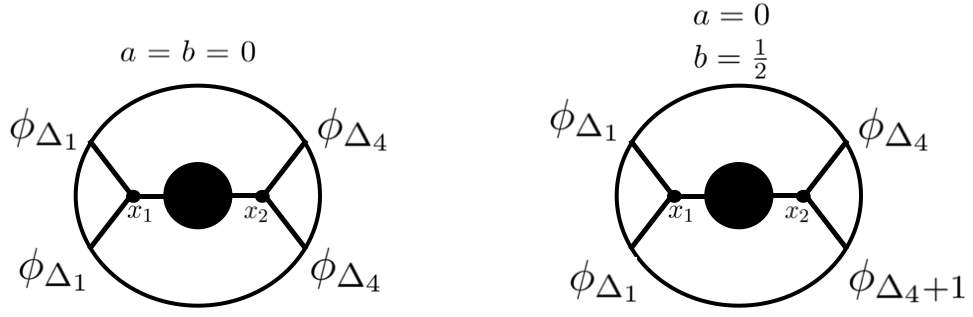}
\caption{The black blob is a general bulk 2-point correlator. \textbf{Left:} The 4-point function with $a=b=0$. \textbf{Right:} The 4-point function with $a=0$, $b=\frac{1}{2}$, discussed in section~\ref{sec:5}. }
\label{fig:figd12}
\end{figure}

In this section we consider scalar 4-point bubble diagrams in $AdS_5$. We generalize the computations of \cite{Carmi:2019ocp} from $AdS_3$ to $AdS_5$. Consider the 4-point function in $d=4$ with external scaling dimensions $a\equiv \frac{\D_2-\D_1}{2} =0$ and $b\equiv \frac{\D_3-\D_4}{2} =\frac{1}{2}$, Fig.~\ref{fig:figd12}-Right. 
A simplification arises from the fact that the $d=4$ conformal block with $a=0$ and $b=\frac{1}{2}$ simplifies from a ${}_2 F_1$ to a power law:
\begin{align} 
\label{eq:cnns7}
&\mathcal{K}^{\D_i}_{\b }(z,\bar z)   = \frac{z\bar z}{z-\bar z} \Big(  z^{\frac{\b}{2}} {}_2 F_1 (\frac{\b}{2},\frac{\b+1}{2},\b,z) \Big) \times \Big( \bar{z}^{\frac{\b}{2}-1}  {}_2 F_1 (\frac{\b-2}{2},\frac{\b-1}{2},\b-2, \bar z)\Big)  -(z \leftrightarrow \bar z)
\nn
&=  \frac{z\bar z}{z-\bar z}\sqrt{\frac{u}{v}}     \frac{2 \sqrt{z}}{1+\sqrt{1-z}}\big(4Z \big)^{-2}\big(4Z \big)^{\b-1}  -(z \leftrightarrow \bar z) \equiv \widetilde{Z}\sqrt{\frac{u}{v}}\big(4Z \big)^{\b-1}
\end{align} 
where the cross-ratios are $u=z\bar z$ and $v=(1-z)(1-\bar z)$, and in the last line we defined:
\begin{align}
Z\equiv \frac{\sqrt{z \bar z}}{(1+\sqrt{1-z})(1+\sqrt{1-\bar z})} \ \ \ , \ \ \ \widetilde{Z}\equiv   \frac{2 \sqrt{z}}{1+\sqrt{1-z}}\big(4Z \big)^{-2} \frac{z\bar z}{z-\bar z} -(z \leftrightarrow \bar z)
\end{align}
The spectral representation of a 4-point function (see e.g Eq.~\ref{eq:fermg4}) simplifies when plugging $\D_2=\D_1$ and $\D_3=\D_4+1$:
\begin{align} 
 g_{4}(z,\bar z ) = \pi 2^{3-d} \int d\nu \tilde{F}_\n \frac{\Gamma_{\frac{d}{2}+i\n -1}  }{4^{i\n} \Gamma_{i\nu}  } ( \G_{\D_1-\frac{d}{4}+\frac{i \nu}{2}} \G_{\D_1-\frac{d}{4}-\frac{i \nu}{2}} \G_{\D_4+\frac{1}{2}-\frac{d}{4}+\frac{i \nu}{2}} \G_{\D_4+\frac{1}{2}-\frac{d}{4}-\frac{i \nu}{2}})  \mathcal{K}^{\D_i}_{\frac{d}{2}+i\n}
\end{align}
Putting $d=4$, and using Eq.~\ref{eq:cnns7} gives:
\begin{align} 
\label{eq:raquel}
g_{4}(z,\bar z )= 2\pi \sqrt{\frac{u}{v}} Z \widetilde{Z} \int_{-\infty}^\infty d\nu\ i \nu \tilde{F}_\n( \G_{\D_1-1+\frac{i \nu}{2}} \G_{\D_1-1-\frac{i \nu}{2}} \G_{\D_4-\frac{1}{2} +\frac{i \nu}{2}} \G_{\D_4-\frac{1}{2} -\frac{i \nu}{2}})  Z^{i\n}
\end{align}
Rewriting this using a gamma function identity, $\G_x \G_{1-x}=\frac{\pi}{\sin (\pi x)}$, gives:
\begin{align} 
\label{eq:poldhhd}
g_{4}(z,\bar z ) =  \sqrt{\frac{u}{v}}  Z \widetilde{Z} \int  d\nu  \frac{ 2\pi^3 \tilde{F}_\n   }{\sin \pi (\D_1-1-\frac{i \nu}{2}) \sin \pi (\D_4-\frac{1}{2}-\frac{i \nu}{2})}   \frac{\G_{\D_4-\frac{1}{2}+\frac{i \nu}{2}}  }{\G_{-\D_4+\frac{3}{2}+\frac{i \nu}{2}}}  \frac{ \G_{\D_1-1 +\frac{i \nu}{2}}  }{\G_{2-\D_1 +\frac{i \nu}{2}}} Z^{i\n}  
\end{align}
Now we close the contour and pick up poles from $ \G_{\D_1-1-\frac{i \nu}{2}}$ at     $i \nu = 2\D_1-2+2m $ , and poles from $ \G_{\D_4-\frac{1}{2}-\frac{i \nu}{2}}$ at     $i \nu = 2\D_4-1+2m$:
\begin{align} 
\label{eq:pldnnd}
&g_{4}(z,\bar z ) \ni  \frac{8\pi^3 \sqrt{\frac{u}{v}}Z \widetilde{Z} }{\cos \pi (\D_1-\D_4 )} \sum_{m=0}^\infty   \frac{\G_{\D_1+\D_4-\frac{3}{2} +m }}{\G_{\D_4-\D_1+\frac{3}{2} +m}} \frac{\G_{2\D_4-1+m} }{\G_{m+1}} \tilde{F}_{ 2\D_4-1+2m}   Z^{ 2\D_4-1+2m} 
\nn
&+ \Big( \D_1 \leftrightarrow \D_4+\frac{1}{2} \Big)
\end{align}
These are contributions from the double-trace poles. As we will see, there can also be poles coming from $\tilde{F}_\n$.

\subsection*{Double-discontinuity}
We can also take the double-discontinuity of the 4-point correlator in Eq.~\ref{eq:poldhhd}, which cancels the sine factors in the denominators\footnote{Where we use
\begin{align}
\label{eq:ddisc}
dDisc_s \big[ \mathcal{K}^{\D_i}_{\frac{d}{2}+i\nu} (z,\bar z) \big] 
=\sin \pi \big(\frac{\Delta_1+\D_2}{2} -\frac{i \nu+\frac{d}{2}}{2}\big) \sin \pi \big(\frac{\Delta_3+\D_4}{2}  -\frac{i \nu+\frac{d}{2}}{2}\big) \mathcal{K}^{\D_i}_{\frac{d}{2}+i\nu} (z,\bar z)
\end{align}}: 
\begin{align} 
\label{eq:poldhhd23}
dDisc [g_{4}(z,\bar z )] =   2\pi^3 \sqrt{\frac{u}{v}}  Z \widetilde{Z} \int  d\nu   \tilde{F}_\n \frac{\G_{\D_4-\frac{1}{2}+\frac{i \nu}{2}}  }{\G_{-\D_4+\frac{3}{2}+\frac{i \nu}{2}}}  \frac{ \G_{\D_1-1 +\frac{i \nu}{2}}  }{\G_{2-\D_1 +\frac{i \nu}{2}}} Z^{i\n}  
\end{align}
The double-discontinuity cancels the double-trace poles, and we are left just with the poles of the internal diagram (i.e poles of $\tilde{F}_\n$). From the double-discontinuity, one can extract the full 4-point correlator by using the conformal dispersion relation \cite{Carmi:2019cub}. Alternatively, one can extract the conformal data by plugging dDisc in the Lorenzian inversion formula \cite{Caron-Huot:2017vep}.

\subsection*{Contact diagrams}
\label{sec:contdi}
The contact diagram has $\tilde{F}_\n =1$, therefore Eq.~\ref{eq:pldnnd} gives:
\begin{align} 
g_{4} (z,\bar z )= \frac{8\pi^3 \sqrt{\frac{u}{v}}Z \widetilde{Z}}{\cos \pi (\D_1-\D_4 )} \sum_{m=0}^\infty      \frac{\G_{\D_1+\D_4-\frac{3}{2} +m }}{ \G_{\D_4-\D_1+\frac{3}{2} +m}} \frac{\G_{2\D_4-1+m} }{\G_{m+1}}  Z^{2\D_4-1+2m}  + \Big( \D_1 \leftrightarrow \D_4+\frac{1}{2} \Big)
\end{align}
This sum gives ${}_2 F_1$'s:
\begin{align}
\label{eq:789s1}
&g_{4}(z,\bar z ) =  \sqrt{\frac{u}{v}}\frac{ \G_{2\D_4-1}  \G_{\D_1+\D_4-\frac{3}{2}}  \widetilde{Z} Z^{2 \D_4}}{(8\pi^3)^{-1}\cos \pi (\D_1-\D_4)} 
\ {}_2 F^{(reg)}_1 (2\D_4-1, \D_1+\D_4-\frac{3}{2},\D_4-\D_1+\frac{3}{2}, Z^{2})
\nn
&+ \Big( \D_1 \leftrightarrow \D_4+\frac{1}{2} \Big)
\end{align}
where ${}_2 F^{(reg)}_1(a,b,c,x) \equiv \frac{1}{\G_c}{}_2 F_1(a,b,c,x)$ is the regularized hypergeometric function.

 \subsection*{Exchange diagrams}
Consider the exchange diagram with internal scaling dimension $\D$, we plug $\tilde{F}_\n = \frac{1}{\n^2+(\D-2)^2}$ in Eq.~\ref{eq:raquel}:
\begin{align} 
g_{4}(z,\bar z )= 2\pi  \sqrt{\frac{u}{v}} Z \widetilde{Z}  \int_{-\infty}^\infty d\n  \frac{ \G_{\D_1-1+\frac{i \nu}{2}} \G_{\D_1-1-\frac{i \nu}{2}} \G_{\D_4-\frac{1}{2} +\frac{i \nu}{2}} \G_{\D_4-\frac{1}{2} -\frac{i \nu}{2}}  }{\n^2+(\D-2)^2}  Z^{i\n}
\end{align}
One pole comes from the exchange operator at $i\n=\D-2$, and there is also a tower of double-trace poles:
\begin{align} 
g_4(z,\bar z )=  g_{exc}(z,\bar z )+g_{d.t} (z,\bar z )
\end{align} 
where:
\begin{align} 
g_{exc}(z,\bar z )=  4\pi^2 \sqrt{\frac{u}{v}}  \widetilde{Z} \frac{1}{2(\D-2)} ( \G_{\D_1-2+\frac{\D}{2}} \G_{\D_1-\frac{\D}{2}} \G_{\D_4 -\frac{3}{2} +\frac{\D}{2}} \G_{\D_4 +\frac{1}{2}-\frac{\D}{2}})  Z^{\D-1}
\end{align}
The contribution from the double-trace poles is:
\begin{align} 
\label{eq:slkjd3}
g_{d.t}(z,\bar z ) =\sqrt{\frac{u}{v}}\frac{8\pi^3   \widetilde{Z}}{\cos \pi (\D_1-\D_4 )} 
 \sum_{m=0}^\infty    \frac{   \frac{\G_{\D_1+\D_4-\frac{3}{2} +m }}{ \G_{\D_4-\D_1+\frac{3}{2} +m}} \frac{\G_{2\D_4-1+m} }{\G_{m+1}}  Z^{2\D_4+2m} }{-(2\D_4-1+2m)^2+(\D-2)^2} + \Big( \D_1 \leftrightarrow \D_4+\frac{1}{2} \Big)
\end{align}
The result of the sum is:
\begin{align} 
&g_{d.t}(z,\bar z ) =  8\pi^3  \sqrt{\frac{u}{v}} \widetilde{Z} \frac{\G_{2\D_4-1}\G_{\D_1+\D_4-\frac{3}{2}}Z^{2\D_4}}{4(\D-2)\cos \pi(\D_1-\D_4)}
 \nn
&\times \bigg[  \G_{\D_4+\frac{\D}{2}-\frac{3}{2}}{}_3 F_2^{(reg)} (2\D_4-1,\D_4+\frac{\D}{2}-\frac{3}{2},\D_1+\D_4-\frac{3}{2}:\D_4+\frac{\D}{2}-\frac{1}{2},\frac{3}{2}+\D_4-\D_1,Z^{2})
\nn
& -\G_{\D_4-\frac{\D}{2}+\frac{1}{2}}{}_3 F_2^{(reg)} (2\D_4-1,\D_4-\frac{\D}{2}+\frac{1}{2},\D_1+\D_4-\frac{3}{2}:\frac{3}{2}-\frac{\D}{2}+\D_4,\frac{3}{2}+\D_4-\D_1,Z^{2}) \bigg]
\nn
&+ \Big( \D_1 \leftrightarrow \D_4+\frac{1}{2} \Big)
\end{align}
 
\subsection*{Loop diagrams}

One can continue and compute loop bubble diagrams in $d=4$, just like we did in the case of $d=2$, \cite{Carmi:2019ocp}. For example, one can use the regularized bubble with $\D=2$:
\begin{align}
\tilde{B}_{reg.}(\n)= \frac{\pi}{4}  \frac{\n^3 \coth(\frac{\pi \n}{2})}{1+\n^2} = \frac{\pi}{4}  \frac{(h-2)^3\cot \frac{\pi h}{2}}{(h-1)(h-3)}  
\end{align}
See Eq.~\ref{eq:pldnnndd8}. For a diagram composed of a sequence of $M$ bubbles, we can plug $\tilde F_\n = (\tilde{B}(\n))^M$ in Eq.~\ref{eq:raquel}:
 \begin{align} 
g_{4}(z,\bar z )= 2\pi \sqrt{\frac{u}{v}} Z \widetilde{Z} \int  d\nu\ i \nu (\tilde{B}_{reg.}(\n))^M ( \G_{\D_1-1+\frac{i \nu}{2}} \G_{\D_1-1-\frac{i \nu}{2}} \G_{\D_4-\frac{1}{2} +\frac{i \nu}{2}} \G_{\D_4-\frac{1}{2} -\frac{i \nu}{2}})  Z^{i\n}
\end{align}
Closing the contour, the sums can be computed in terms of hypergeometric functions.


\section{Scalar 4-point bubble diagrams in $AdS_2$}
\label{sec:E2}

Consider the 4-point function of scalar bubble diagrams in $AdS_2$ with equal external and internal scaling dimensions $\D$. We start by writing the spectral representation of the diagram in Fig.~\ref{fig:figd12}-Left:
\begin{align}
\label{eq:1293} 
g_4(z )=  \int  d\n  \tilde{F}(\n) \frac{\Gamma_{\Delta -\frac{d+2i\nu}{4}}^2\Gamma_{\Delta -\frac{d-2i\nu}{4}}^2\Gamma_{\frac{d+2i\nu}{4}}^4}{\Gamma_{i\nu}\Gamma_{\frac{d}{2}+i\nu}}\mathcal{K}_{\frac{d}{2}+i\nu}(z ) \,,
\end{align}

The conformal block in $d=1$ is:
\begin{align}
\mm{K}_\b^{(d=1)}(z) = z^\b {}_ 2F_1 (\b,\b,2\b,z) = \frac{2\Gamma_{2\b}}{\Gamma^2_{\b}}Q_{\b-1}(\hat z)
\end{align}


The poles of the 1-loop bubble in $d=1$ are:
\begin{align}
 2\tilde{B}(\nu)\overset{\frac{1}{2}+i\nu\, \sim  \,2\Delta+2n}{\sim}-\frac{1}{\frac{1}{2}+i\nu-(2\Delta+2n)}\frac{ \G_{n+\frac{1}{2}}\Gamma^2_{\Delta+n} \Gamma_{2\Delta+n-\frac{1}{2}}}{\sqrt \pi \Gamma_{n+1} \G^2_{\Delta+n+\frac{1}{2}}\Gamma_{2\Delta+n}}\,.
 \end{align}

The spectral representation for the $M$ bubble diagram is $\tilde{F}(\n) = \big( \tilde{B}(\nu) \big)^M$, so we have in $d=1$,:
\begin{align} 
\label{eq:dfmmfff}
g_4(z ) =
\sum_{n=0}^\infty  \frac{d^{M+1}}{dn^{M+1}}\bigg(\bigg(  \frac{\G_{n+\frac{1}{2}}\Gamma^2_{\Delta+n} \Gamma_{2\Delta+n-\frac{1}{2}}}{\sqrt \pi \Gamma_{n+1} \G^2_{\Delta+n+\frac{1}{2}}\Gamma_{2\Delta+n}}
\bigg)^M
\frac{\G^2_{\D+n}\G^2_{2\D+n-\frac{1}{2}}}{\G^2_{n+1}\G^2_{\D+n+\frac{1}{2}}} (4\D+4n-1) Q_{2\D+2n-1} (\hat{z})  \bigg)  
\end{align}


For simplicity let's consider $\D=1$. The contact diagram $M=0$ gives:
\begin{align}
g_4(z )= 4\sum_{n=0}^\infty \frac{d}{dn} \bigg(  (4n+3) Q_{2n+1} (\hat{z}) \bigg)
\end{align}
The result of this sum gives:
\begin{align}
g_4(z )=  2z^2 \Big(\frac{\log(1-z)}{z} +\frac{\log(z)}{1-z} \Big) 
\end{align}
which precisely matches Eq~7.29 of \cite{Mazac:2018ycv}. 

One can also compute the 1-loop bubble, i.e $M=1$ and $\D=1$. From Eq.~\ref{eq:dfmmfff} we have:
\begin{align} 
g_4(z ) =
\sum_{n=0}^\infty  \frac{d^{2}}{dn^{2}}\bigg(\frac{4n+3}{(2n+2)(2n+1)} Q_{2n+1} (\hat{z})  \bigg) 
\end{align}
Using Eq.~\ref{eq:pdpderp}, we can also write this as follows:
\begin{align} 
\label{eq:fdn99f}
w_z g_4(z ) =
\sum_{n=0}^\infty  \frac{d^{2}}{dn^{2}}\bigg( (4n+3) Q_{2n+1} (\hat{z})  \bigg) 
\end{align}
where the differential operator is defined as $w_z \equiv -\frac{d}{d\hat{z}} (1-\hat{z}^2)\frac{d}{d\hat{z}}$.
The RHS of Eq.~\ref{eq:fdn99f} has a canonical form, and it would be interesting to see if it can be computed e.g from Eq.~5.9 of \cite{Carmi:2019ocp}. It would be interesting to compare the result with that of Eq.~7.34 of \cite{Mazac:2018ycv}.


\bibliographystyle{utphys}
\bibliography{bulkdiagrams}

\end{document}